\documentclass[pra,aps,twocolumn,superscriptaddress]{revtex4-1}
\usepackage{amsfonts}
\usepackage{epsfig,amsmath}
\usepackage{subfigure}
\usepackage{graphicx}
\usepackage{dcolumn}
\usepackage{stmaryrd}
\usepackage{mathrsfs}
\usepackage{pifont}
\usepackage{amsthm}
\usepackage{braket}
\usepackage{amssymb}
\usepackage{bm}
\usepackage{latexsym}
\usepackage[colorlinks=true,linkcolor=blue,citecolor=blue]{hyperref}
\usepackage{color}
\usepackage{epstopdf}
\usepackage[ruled]{algorithm2e}

\begin{document}

\title{Precision limit under weak coupling with an ancillary qubit}

\author{Peng Chen}
\affiliation{School of Physics, Zhejiang University, Hangzhou 310027, Zhejiang, China}

\author{Jun Jing}
\email{Contact author: jingjun@zju.edu.cn}
\affiliation{School of Physics, Zhejiang University, Hangzhou 310027, Zhejiang, China}

\begin{abstract}
We propose a measurement-based quantum metrology protocol in a composite model, where the probe system (a spin ensemble) is coupled to an ancillary two-level system (qubit) with a general Heisenberg XXZ interaction. With optimized probe-ancilla coupling strengths and duration of joint evolution, the two parallel evolution paths of the probe system induced by the unconditional measurement on qubit can transform an eigenstate of the collective angular momentum operator of spin ensemble into a two-component state with a large distance in eigenspace. The quantum Fisher information about the phase encoded in the probe system of polarized states or their superposition, that could be relaxed to mixed states, can therefore manifest an exact or asymptotic quadratic scaling with respect to the probe size (spin number) $N$. The quadratic scaling behavior is found to be insensitive to the imperfect encoding operator, polarized direction of probe, and coupling strength. The phase sensitivity can approach the Heisenberg limit by virtue of the parity detection on either ancillary qubit or probe system. This work justifies that the unconditional measurement on a weakly coupled qubit could be an efficient resource to replace Greenberger-Horne-Zeilinger-type states and squeezing Hamiltonian for exceeding the standard quantum limit in metrology precision.
\end{abstract}

\maketitle

\section{Introduction}

As an emerging field that integrates quantum mechanics with statistics, quantum metrology~\cite{giovannetti2004quantum,Sun2010fisher,ma2011quantum,genoni2012optical,escher2012quantum,zhong2013fisher} holds significant importance for a variety of quantum sciences and technologies, including atomic clock~\cite{Katori2011optical,Ludlow2015optical}, gravitational wave detection~\cite{Caves1981quantum}, biological sensing~\cite{Taylor2016quantum,Mauranyapin2017evanescent}, and magnetometry~\cite{Jones2009magnetic}. The general goal of quantum metrology is to achieve a high sensitivity in parameter estimation. A conventional quantum estimation protocol~\cite{Sun2010fisher,ma2011quantum,escher2012quantum} consists of (1) preparing the probe into a resource state, (2) encoding the to-be-estimated parameter into the probe state, (3) performing measurements or detections on the probe to extract parametric information, and (4) processing the outcomes to estimate the parametric precision. The metrology precision using uncorrelated probe states is bounded by the standard quantum limit (SQL), which scales as $1/\sqrt{N}$ with $N$ the subunit number or the measurement number~\cite{kay1993fundamentals}. The quantum mechanics in metrology~\cite{giovannetti2006quantum} is highlighted with the scaling advantage of the Heisenberg limit (HL) over SQL. It depends on resource states and presents a scaling law inverse to $N$.

A common strategy to enhance metrology precision is the use of certain nonclassical states as resource states, including the many-body entangled states and squeezed spin states. Highly entangled states have been prepared and the Heisenberg limit is observed in small-scale systems~\cite{song2019generation,chalopin2018quantum,kaufmann2017scalable}. A quantum processor of $18$ qubits has been prepared as a Greenberger-Horne-Zeilinger (GHZ) state with a fidelity $\sim0.525$. However, the generation of GHZ states for a large atomic ensemble remains a significant challenge due to the high susceptibility to environmental decoherence. Squeezed spin states~\cite{kitagawa1993squeezed,wineland1994squeezed,ma2011quantum} are another widely explored class of nonclassical resources, which enhance the measurement sensitivity through suppressing quantum fluctuations. Also they are featured with a higher tolerance to decoherence and a lower loss than GHZ states~\cite{zhang2013quantum}. Collective one-axis twisting (OAT) Hamiltonian $H_{\rm OAT}=\chi J_z^2$ is widely applied to deterministically produce strong squeezing, which yields a sub-HL noise-reduction $\propto1/N^{2/3}$ for $N$ particles with the optimized duration $\chi t\simeq3^{1/6}N^{-2/3}$~\cite{wineland1992spin,kitagawa1993squeezed}. The nonlinear atomic collisions in two-component Bose-Einstein condensate give rise to an OAT interaction~\cite{sorensen2001many,gross2010nonlinear,riedel2010atom}. In cavity quantum electrodynamics, the OAT Hamiltonian can be engineered via dispersive atom-light interaction~\cite{schleier2010squeezing,leroux2010implementation,norcia2018cavity}, in which the atom-atom interaction is mediated by photons within the cavity. Other studies to produce OAT interaction can be found in trapped ions~\cite{molmer1999multiparticle,britton2012engineered,bohnet2016quantum} and lattice systems~\cite{sorensen1999spin,he2019engineering,hernandez2022one}.

Exploiting non-entangled and non-squeezed probes for quantum metrology has gained considerable attention in recent years, exemplified by the coupling or the entanglement between the probe and an ancillary system as an alternative resource for outperforming SQL in the parametric estimation~\cite{boixo2007generalized,xia2023nanoradian,yang2022variational,luo2023time,fan2024achieving,chen2024qubit} or promoting robustness against noise~\cite{demkowicz2014using,he2021quantum}. In a protocol for measuring the frequency of probe units~\cite{fan2024achieving}, the parameter information can be efficiently extracted via measurements on the ancillary qubit, after tracing out the probe system. By properly tailoring the interaction Hamiltonian, the time points for measurements, and the coupling strength, the estimation precision can periodically reach the Heisenberg scaling in terms of probe size and time.

The quantum metrology protocols assisted by the external interaction~\cite{yang2022variational,luo2023time,fan2024achieving,chen2024qubit}, however, demand the strong coupling strength, specific interaction Hamiltonian, or multi-outcome measurements. Under the time-reversal strategy~\cite{pezze2017optimal,yurke19862,chen2024qubit}, projective measurement can be scheduled to saturate quantum Cram\'{e}r-Rao bound for parameter estimation, given an encoded probe state. In a phase estimation~\cite{luo2023time}, the interaction between a photonic system and an ancillary qubit can be controlled by an external field, enabling unitary transformation along the forward- and reverse-time directions. The interaction generates a photonic entangled state to improve the quantum Fisher information and the time-reversal strategy allows the classical Fisher information to saturate its quantum counterpart. Yet the transversal coupling strength is required to be tunable and comparable in magnitude to the qubit frequency. In a protocol for a spin-ensemble probe, the time-reversal strategy can be realized with an ancillary qubit~\cite{chen2024qubit}. However, the ZZ interaction in that protocol remains a challenge to implement in experimental platforms. Moreover, the unconditional measurement on ancillary qubit can become an unconventional resource that replaces large-scale entanglement in standard quantum metrology protocols~\cite{ChenJing2025}. Irrelevant of the idle evolution after the parametric encoding, the quantum Fisher information can be saturated by multi-outcome projective measurements on the probe system. A general interaction Hamiltonian, a less-constraint ancillary qubit, and a more efficient measurement strategy are therefore highly desired in a qubit-assisted metrology protocol, which is the motivation of the current work.

In this paper, we propose a high-precision protocol based on a general Heisenberg XXZ interaction between the spin-ensemble probe and an ancillary qubit, which significantly expands the application range of the previous protocols involving merely specific interaction. Also it does not require a strong or time-varying coupling strength. The unconditional measurement on qubit can induce two parallel evolution paths of the probe system. When the strengths of the longitudinal and transversal interactions are close to each other, an eigenstate of the collective angular momentum operator of the probe system can then be transformed to a superposed state distributed with a large distance in eigenspace. If the probe is prepared as a polarized state along an optional optimized direction, it can evolve to a GHZ-type state after the unconditional measurement on qubit. An asymptotic Heisenberg-scaling behavior about the metrology precision can be obtained even when the probe is prepared as a thermal state. This scaling behavior is insensitive to the imprecise phase encoding and probe preparation and the fluctuated coupling strengths of both transversal and longitudinal directions. To efficiently extract information about the to-be-estimated parameter, we perform the parity detections on either ancillary qubit or probe, both of which approach HL in the phase sensitivity.

The rest of this work is structured as follows. Section~\ref{metrologyProtocol} introduces the circuit model of our metrology protocol. In Sec.~\ref{QFIofMetrology}, we study the conditions about the systematic parameters and the initial state of probe for attaining the Heisenberg limit. In Sec.~\ref{imprecisecontrol}, we discuss the sensitivity of our metrology protocol to imperfect probe state preparation and imprecise controls over the encoding direction and the coupling strengths. We analyze the phase sensitivity under parity detection performed on the ancillary qubit and the probe system in Secs.~\ref{detection on the qubit} and \ref{detection on the probe}, respectively. The entire work is summarized in Sec.~\ref{conclusion}. Appendix~\ref{app} provides a detailed derivation of the joint evolution operator for the full Hamiltonian.

\section{Model and Hamiltonian}\label{metrologyProtocol}

Consider a quantum metrology model about a large-spin probe (spin ensemble) coupled to an ancillary spin-$1/2$ (qubit) by a general Heisenberg XXZ interaction. The full Hamiltonian can be written as ($\hbar\equiv1$)
\begin{equation}\label{H}
H=\omega_PJ_z+\omega_A\sigma_z+g_zJ_z\sigma_z+g(J_x\sigma_x+J_y\sigma_y),
\end{equation}
where $J_{\mu}=\sum_{l=1}^N\sigma_\mu^l/2$, $\mu=x,y,z$, represents the collective spin operator with $N$ the total spin number of the probe ensemble and $\sigma_\mu^l$ the $\mu$-component of Pauli operator for the $l$th probe spin. $\sigma_{\mu}$ is the Pauli matrix of the ancillary qubit. $\omega_P$ and $\omega_A$ denote the energy splitting of the probe spin and the ancillary spin, respectively. $g$ and $g_z$ denote the transversal and longitudinal coupling strengths, respectively.

\begin{figure}[htbp]
\begin{centering}
\includegraphics[width=0.9\linewidth]{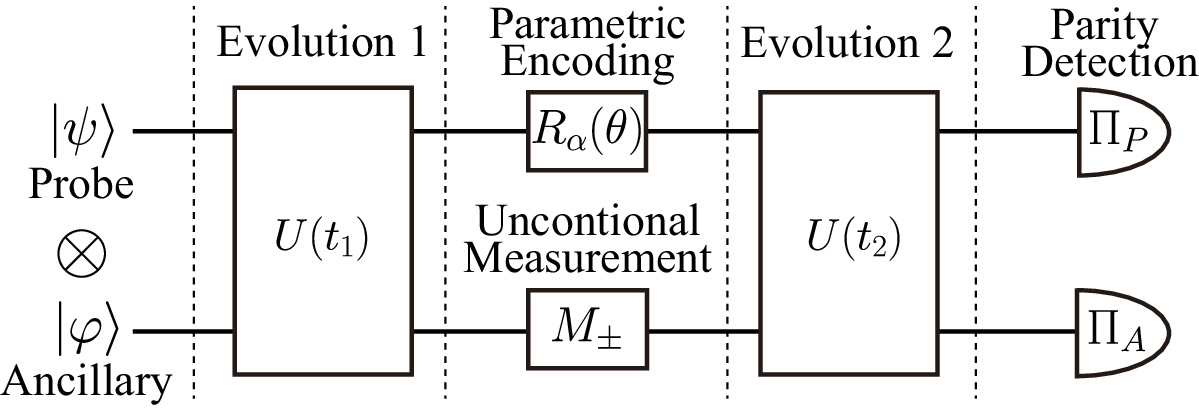}
\caption{A circuit model of our measurement-based metrology. The composite system prepared in a separable state $|\psi\rangle\otimes|\varphi\rangle$ experiences two stages of free joint unitary evolutions $U(t_1)$ and $U(t_2)$. In between them, a to-be-estimated phase parameter $\theta$ is encoded into the large-spin-probe system via a unitary rotation $R_{\alpha}(\theta)$ and meanwhile an unconditional measurement $M_{\pm}=|\pm\rangle\langle\pm|$ in the basis of $\sigma_x$ is performed on the ancillary qubit. The output is determined by the parity detections on the probe system $\Pi_P$ or the ancillary qubit $\Pi_A$.}\label{protocol}
\end{centering}
\end{figure}

The Heisenberg XXZ interaction in Eq.~(\ref{H}) is popular in various experimental platforms. In nitrogen-vacancy (NV) center systems~\cite{barry2020sensitivity}, $J_\mu$ describes the $^{14}N$ nuclear spins and $\sigma_\mu$ describes the NV electron spin consisting of levels $m_s=0,1$ that are addressed by a magnetic field parallel to the quantization axis of the NV center. Under the external magnetic field $B=10^{3}$ G, the probe-spin frequency is about $4.63$ MHz, the ancillary-spin frequency is about $2.87$ GHz, and their coupling strengths $g_z$ and $g$ are found to be about $2.30$ MHz and $2.10$ MHz~\cite{he1993paramagnetic}, respectively. In quantum-dot systems~\cite{yao2006theory,liu2007control}, $J_\mu$ and $\sigma_\mu$ describe the nuclear spins and the electron spin, respectively. The probe frequency is about $\omega_P=5$ MHz, the ancillary frequency is about $\omega_A=2.5$ GHz, and the coupling strengths $g_z\approx g\approx 80$ MHz, under the external magnetic field $B=379$ G. It turns out that (i) the coupling strengths are much weaker than the system frequency and (ii) the longitudinal coupling strength is usually close to the transversal one. Our metrology method can be practiced in these systems, since the Heisenberg scaling in metrology precision can be achieved as long as $\sqrt{g_z^2-g^2}=(2\omega_A-\omega_P)/(N+1)$ [see Eq.~(\ref{optimal g gz})].

As shown in Fig.~\ref{protocol}, the probe spin ensemble and the ancillary qubit are assumed to be initially separable, i.e., the input state of the composite system is a product state $|\psi\rangle\otimes|\varphi\rangle$. The {\em nonunitary} evolution operator for the entire circuit can be written as
\begin{equation}\label{U}
U_{\theta,\pm}=U(t_2)R_\alpha(\theta)M_\pm U(t_1)=e^{-iHt_2}e^{-i\theta J_\alpha}|\pm\rangle\langle\pm|e^{-iHt_1},
\end{equation}
where $t_1$ and $t_2$ are the durations of the two stages of joint unitary evolution, and $|\pm\rangle\equiv(|e\rangle\pm|g\rangle)/\sqrt{2}$ are the eigenbases of $\sigma_x$ with $|g\rangle$ and $|e\rangle$ being the ground and excited states of the ancillary qubit. The spin rotation $R_\alpha(\theta)=\exp(-i\theta J_\alpha)$ describes the parametric encoding into the probe, where $J_\alpha=\cos\alpha J_x+\sin\alpha J_y$ is a general collective angular momentum operator lying in the equatorial plane of the Bloch sphere, oriented at an arbitrary angle $\alpha$ with respect to the $x$ axis. The rotation around an arbitrary direction can be practiced by a sequence of rotations $R_\alpha(\theta)=R_z(\alpha)R_y(\pi/2)R_z(\theta)R_y(-\pi/2)R_z(-\alpha)$, where $R_z(\pm\alpha)$ and $R_y(\pm\pi/2)$ indicate $\alpha$ and $\pi/2$ pulses applied along the $z$ and $y$ axes, respectively, and $R_z(\theta)$ could be generated by a dispersive coupling between the probe system and a to-be-measured system~\cite{meyer2001experimental,gross2010nonlinear,ockeloen2013quantum}. Without loss of generality, we assume $J_\alpha=J_x$, such that $R_\alpha(\theta)=R_x(\theta)$ in Eq.~(\ref{U}). The projection operator $M_\pm\equiv|\pm\rangle\langle\pm|$ describes the unconditional measurement on the ancillary qubit, where ``unconditional'' means the whole process continues irrespective to the measurement outcome that would not be recorded. In experiments, the measurement over the ancillary qubit, e.g., the electron spin in NV center, can be realized by the collected fluorescence~\cite{xie2021beating}. At the end of the entire circuit, a parity detection will be performed on either ancillary qubit or probe system (see Sec.~\ref{Phasesensitivity}). The output signal and fluctuation can be used to infer the phase sensitivity about the unknown $\theta$.

\section{Quantum Fisher information of the measurement-based metrology}\label{QFIofMetrology}

Previous studies~\cite{giovannetti2006quantum,pang2014quantum} show that the quantum Fisher information associated with the phase to be estimated can be written as the variance of a dynamical generator with respect to the probe state. The maximum QFI is attained when the probe is prepared in an equal superposition of the eigenstates with the maximum and minimum eigenvalues of the phase generator, such as the GHZ state in atomic systems or the NOON state in photonic systems. For a phase generator $J_x$ of our large-spin system, the relevant GHZ-type state takes the form
\begin{equation}\label{GHZ}
\begin{aligned}
|\mathrm{GHZ}\rangle&=\frac{1}{\sqrt{2}}\left(|j,\pm j\rangle_x+e^{-i\phi}|j,\mp j\rangle_x\right)\\
&=\frac{1}{\sqrt{2}}\left(\mathcal{I}^{N+1}+e^{-i\phi}e^{-i\pi J_z}\right)|j,\pm j\rangle_x,
\end{aligned}
\end{equation}
where $j=N/2$ is the quantum number with $N$ the size of the probe spin ensemble, $\mathcal{I}^{N+1}$ is the identity matrix of $N+1$ dimensions, $\phi$ is a local phase, and $|j,m\rangle_x$ is the eigenstate of the collective spin operators $J_x$ with eigenvalue $m$ for $-j\le m\le j$. Here we have applied $|j,\mp j\rangle_x=e^{-i\pi J_z}|j,\pm j\rangle_x$ in the second line. GHZ state can be prepared by allowing the polarized state $|j,\pm j\rangle_x$ to undergo a two-path evolution, i.e., $\mathcal{I}^{N+1}$ and $e^{-i\pi J_z}$, a possible strategy for which is indefinite gate order or dynamics~\cite{zhao2020quantum,yin2023experimental,xia2023nanoradian}. For example, in a protocol about estimating the expectation-value product of position and momentum displacements of a continuous-variable system~\cite{zhao2020quantum}, an ancillary two-level system acts as a quantum switch, creating a superposition of two paths. Along one path, all the position-displacement operators are performed before all the momentum-displacement operators; and along another one, the order is inverted. The estimation precision can therefore be enhanced up to the super-Heisenberg scaling in terms of the repetition number of position or momentum displacements.

In the circuit of Fig.~\ref{protocol}, our protocol employs the free joint unitary evolution for Evolution 1 and the unconditional measurement $M_\pm$ to construct the two-path evolution. Distinct measurement outcome about the ancillary qubit indicates distinct path of output. Despite we obtain a classical mixture of two paths on different outcomes, the final QFI attains HL as long as $N$ is sufficiently large. Suppose the initial state of the qubit is $|\varphi\rangle=|+\rangle$ and according to Eq.~(\ref{U}), the unnormalized output state of the composite system then reads
\begin{equation}\label{unnormalized state}
|\Psi_\pm\rangle=M_\pm U(t_1)|\psi\rangle\otimes|\varphi\rangle=\langle\pm|U(t_1)|+\rangle|\psi\rangle\otimes|\pm\rangle,
\end{equation}
with a probability $\mathcal{N}_\pm=\langle\Psi_\pm|\Psi_\pm\rangle$ after the first-stage evolution $U(t_1)$ and a measurement on qubit in its $\sigma_x$ basis. Here the subscript $\pm$ indicates the measurement outcome. In particular, we have
\begin{equation}\label{unnormalized state PlusPlus}
\begin{aligned}
&|\Psi_+\rangle=\langle+|U(t_1)|+\rangle|\psi\rangle\otimes|+\rangle\\
&=\frac{O_-+e^{-2i\Omega(J_z)t_1}O_+}{4}e^{i[\Omega(J_z)-A(J_z)]t_1}|\psi\rangle\otimes|+\rangle,
\end{aligned}
\end{equation}
with the operator
\begin{equation}
\begin{aligned}
&O_{\pm}=1\pm\frac{\Lambda(J_z)}{\Omega(J_z)}\\
+&e^{i\omega_Pt_1}e^{\pm i[\Omega(J_z)-\Omega(J_z-1)]t_1}\left[1\pm\frac{\Lambda(J_z-1)}{\Omega(J_z-1)}\right]\\
\pm& e^{i\omega_Pt_1}e^{i[\Omega(J_z)-\Omega(J_z+1)]t_1}\frac{g}{\Omega(J_z)}J_-\pm J_+\frac{g}{\Omega(J_z)},
\end{aligned}
\end{equation}
where the operator functions are defined as
\begin{subequations}
\begin{align}
\omega(J_z)&\equiv g\sqrt{(j-J_z)(j+J_z+1)}.\label{function omega}\\
\Lambda(J_z)&\equiv g_z\left( J_z+\frac{1}{2}\right)+\Delta_A,\label{function lambda} \\
\Omega(J_z)&\equiv \sqrt{\omega^2(J_z)+\Lambda^2(J_z)}.\label{function Omega}\\
A(J_z)&\equiv \omega_P\left( J_z+\frac{1}{2}\right)-\frac{g_z}{2}.\label{function A}
\end{align}
\end{subequations}
Here $\Delta_A\equiv\omega_A-\omega_P/2$ is the detuning between the ancillary qubit and the probe spin. The derivation details of the free joint evolution operator $U(t)$ is provided in Appendix~\ref{app}. It is found that when
\begin{subequations}
\begin{align}
&e^{-2i\Omega(J_z)t_1}=e^{-i\phi}e^{-i\pi J_z},\label{sufficient condition Omega}\\
&|\langle j,\pm j|_x(O_-+O_+)e^{i[\Omega(J_z)-A(J_z)]t_1}|\psi\rangle|=4\sqrt{2\mathcal{N_+}},
\label{sufficient condition probe}
\end{align}
\end{subequations}
the probe state $|\Psi_+\rangle$ in Eq.~(\ref{unnormalized state PlusPlus}) becomes the same as the GHZ-type state in Eq.~(\ref{GHZ}) up to a global phase. In this case, $|\Psi_+\rangle$ is a superposition of two parallel evolution paths respectively indicated by $\mathcal{I}^{N+1}$ and $e^{-2i\Omega(J_z)t_1}$.

With Eqs.~(\ref{function omega}), (\ref{function lambda}), and (\ref{function Omega}), the solutions to Eq.~(\ref{sufficient condition Omega}) can be found as
\begin{subequations}
\begin{align}
&\sqrt{g_z^2-g^2}=\frac{2\Delta_A}{N+1},\label{optimal g gz}\\
&t_1=t_{1,{\rm opt}}(n_1)\equiv\frac{(N+1)(2n_1+1)}{4}\frac{\pi}{\Delta_A},\label{optimal time}
\end{align}
\end{subequations}
where $n_1$ is an integer. Equation~(\ref{optimal g gz}) suggests that the transversal coupling strength $g$ in our protocol should not be larger than the longitudinal coupling strength $g_z$. Despite their difference scales inversely with the probe size $N$, the magnitudes of both $g_z$ and $g$ are not relevant to $N$. Moreover, when $g=0$, the constraints in Eqs.~(\ref{optimal g gz}) and (\ref{optimal time}) can be reduced to those for the $ZZ$ interaction between probe and ancillary qubit~\cite{ChenJing2025}. Consequently, the phase parameter $\phi$ in Eq.~(\ref{sufficient condition Omega}) reads,
\begin{equation}\label{parameter phi}
\phi=(2n_1+1)\left[\frac{1}{2}+(N+1)^2\frac{g_z}{4\Delta_A}\right]\pi.
\end{equation}
In the large-number limit of $N$, we have $g\approx g_z$ and hence $\Lambda(J_z)/\Omega(J_z)\approx(2J_z+1)/(N+1)$, $gJ_+/\Omega(J_z)\approx2J_+/(N+1)$, according to Eqs.~(\ref{optimal g gz}), (\ref{function lambda}), and (\ref{function omega}). Subsequently, a proper probe state
\begin{equation}\label{probeOptimal}
|\psi\rangle=|j,\pm j\rangle_{\rm opt}=e^{-i[\Omega(J_z)-A(J_z)]t_{1,{\rm opt}}}|j,\pm j\rangle_x
\end{equation}
and a proper eigenfrequency
\begin{equation}\label{optimalFrequency}
\omega_P=\frac{n_P}{t_{1,{\rm opt}}}\pi
\end{equation}
with $n_P$ integer constitute a solution to Eq.~(\ref{sufficient condition probe}).  Here $|j,m\rangle_{\rm opt}$'s with $-j\leq m\leq j$ denote the eigenstates for the optimized collective angular momentum operator
\begin{equation}\label{Jopt_operator}
\begin{aligned}
J_{\rm opt}&=e^{-i[\Omega(J_z)-A(J_z)]t_{1,{\rm opt}}}J_xe^{i[\Omega(J_z)-A(J_z)]t_{1,{\rm opt}}} \\
&\approx(-1)^{(n_P-n_1)}J_y.
\end{aligned}
\end{equation}

Using one of the optimized probe states in Eq.~(\ref{probeOptimal}), $|\psi\rangle=|j,-j\rangle_{\rm opt}$, and the conditions in Eqs.~(\ref{optimal g gz}), (\ref{optimal time}), and (\ref{optimalFrequency}), the unnormalized state of the composite system upon obtaining $|\pm\rangle$ of the qubit state before the second stage of joint evolution $U(t_2)$ can be respectively expressed as
\begin{equation}\label{output state}
\begin{aligned}
|\Psi_{\theta,\pm}\rangle&\approx\frac{1\mp i(-1)^{n_1+n_P}}{4(N+1)}R_x(\theta)\\
&\left\{\left[(N\mp2J_x)\pm 2i(-1)^{n_1+n_P}J_{\mp,x}\right]\right.\\
&\pm i(-1)^{n_1(N+1)+n_P}e^{-i\phi}e^{-i\pi J_z}\\
&\left.\left[(N+2\mp2J_x)\pm 2i(-1)^{n_1+n_P}J_{\pm,x}\right]\right\}\\
&|j,-j\rangle_x\otimes|\pm\rangle,
\end{aligned}
\end{equation}
where $J_{\pm,x}\equiv-J_z\pm iJ_y$. And the relevant probabilities are unbalanced and read
\begin{equation}\label{probability measurement result}
\mathcal{N}_{\theta,\pm}=\langle\Psi_{\theta,\pm }|\Psi_{\theta,\pm }\rangle =\frac{1}{2}\left(1\pm\frac{N}{N+1}\right).
\end{equation}
When $N\gg1$, it is found that $\mathcal{N}_{\theta,+}\rightarrow1$ and $\mathcal{N}_{\theta,-}\rightarrow0$, indicating that the measurement result about the ancillary qubit is almost definitely $|+\rangle$. For either measurement outcome, the effective QFI is defined as~\cite{braunstein1994statistical}
\begin{equation}\label{QFI pm}
F_{Q,\pm}=4\mathcal{N}_{\theta,\pm}
\left[\langle\partial_\theta\Psi'_{\theta,\pm}|\partial_\theta\Psi'_{\theta,\pm}\rangle
-|\langle\Psi'_{\theta,\pm}|\partial_\theta\Psi'_{\theta,\pm}\rangle|^2\right],
\end{equation}
where the normalized state is $|\Psi'_{\theta,\pm}\rangle=|\Psi_{\theta,\pm}\rangle/\sqrt{\mathcal{N}_{\theta,\pm}}$. Thus, the full QFI reads
\begin{equation}\label{QFI optimal}
F_Q=F_{Q,+}+F_{Q,-}\approx N^2\left[1-\frac{4}{(N+1)^2}\right].
\end{equation}
Then for a large-size probe, i.e., $N\gg1$, the quantum Fisher information can be approximated as $F_Q\approx N^2$.

As for another optimized state $|\psi\rangle=|j,j\rangle_{\rm opt}$ given in Eq.~(\ref{probeOptimal}), one can obtain an unnormalized state similar to Eq.~(\ref{output state}) under the same optimal conditions in Eqs.~(\ref{optimal g gz}), (\ref{optimal time}), and (\ref{optimalFrequency}). The relevant probability in Eq.~(\ref{probability measurement result}) now becomes $\mathcal{N}_{\theta,\pm}=[1\mp N/(N+1)]/2$. It means that the ancillary qubit after the unconditional measurement is almost certainly found in $|-\rangle$ under the large-$N$ limit. Consequently, one can obtain the same result as Eq.~(\ref{QFI optimal}).

The preceding analysis demonstrates that the Heisenberg scaling in metrology precision can be approached when the probe and the ancillary qubit are prepared at a polarized state $|\psi\rangle$ in Eq.~(\ref{probeOptimal}) and $|\varphi\rangle=|+\rangle$, respectively, and the joint evolution time is scheduled as $t_1$ in Eq.~(\ref{optimal time}) before encoding in probe and measurement on qubit. Similarly, if the ancillary qubit is initialized as $|\varphi\rangle=|-\rangle$, one can obtain the same sufficient conditions in Eqs.~(\ref{sufficient condition Omega}) and (\ref{sufficient condition probe}) or Eqs.~(\ref{optimal g gz}), (\ref{optimal time}), (\ref{probeOptimal}), and (\ref{optimalFrequency}) under the large-$N$ limit for generating the GHZ-type state. QFI also shows the same scaling behavior as in Eq.~(\ref{QFI optimal}). In the following, we then always assume $|\varphi\rangle=|+\rangle$ and $|\psi\rangle=|j,-j\rangle_{\rm opt}$ unless stated otherwise.

Next, we consider a more general case, where the probe system is prepared as a pure state superposed by the eigenbases of the specific optimized collective angular momentum operator in Eq.~(\ref{Jopt_operator}), e.g., $|\psi_m\rangle=a_m|j,m\rangle_{\rm opt}+b_me^{-i\phi_m}|j,-m\rangle_{\rm opt}$ with $1\leq m\leq j$ and real numbers $a_m$, $b_m$, and $\phi_m$. Upon the preceding optimal conditions in Eqs.~(\ref{optimal g gz}), (\ref{optimal time}), and (\ref{optimalFrequency}), one can still obtain a square-scaling QFI for $m$:
\begin{equation}\label{QFI general}
\begin{aligned}
F_Q&\approx4\left[1-\frac{6}{(N+1)^2}\right]m^2-\frac{\left[(N+1)^2-4m^2\right]^2}{(N+1)^2-4m^2(a_m^2-b_m^2)^2}\\
&\times\frac{16m^2}{(N+1)^2}a_m^2b_m^2\sin^2\phi_m\sin^2\left(\phi+\frac{N+1}{2}\pi\right)\\
&+(-1)^N8m\frac{N(N+2)-4m^2}{(N+1)^2}a_mb_m\\
&\times\sin\phi_m\sin\left(\phi+\frac{N+1}{2}\pi\right)+\frac{2N^2}{(N+1)^2}
\end{aligned}
\end{equation}
up to the order of $N^0$. Given a quantum number $m$, QFI can be approximated as $F_Q\approx4[1-6/(N+1)^2]m^2+2N^2/(N+1)^2$ under any of the following situations: (i) $a_mb_m=0$, (ii) $\phi_m=n_m\pi$, or (iii) $\phi+(N+1)\pi/2=n_\phi\pi$, with $n_m$ and $n_\phi$ integers. It further reduces to Eq.~(\ref{QFI optimal}) for $m=j$. Situations (i) and (ii) correspond to optimizing the input state, or more precisely, the population distribution and the relative phase $\phi_m$ of the probe system. Situation (iii) depends only on the ratio of the systematic parameters $g_z/\Delta_A$ in Eq.~(\ref{parameter phi}), which is described by
\begin{equation}
g_z=\frac{N+2n_\phi-2}{2n_1+1}\frac{2\Delta_A}{(N+1)^2}.
\end{equation}
Here we have used the optimized joint evolution time $t_{1,{\rm opt}}$ in Eq.~(\ref{optimal time}). Equation~(\ref{QFI general}) therefore confirms that the polarized states $|j,\pm j\rangle_{\rm opt}$ in Eq.~(\ref{probeOptimal}) together with the unconditional measurement on ancillary qubit can be regarded as an efficient resource for approaching HL in metrology precision.

An asymptotic Heisenberg-scaling behavior of QFI can however appear even when the probe state is prepared as a mixed state, e.g., the thermal state in the bases of the optimized collective angular momentum operator $J_{\rm opt}$ in Eq.~(\ref{Jopt_operator}). We then have
\begin{equation}\label{thermal state}
\rho_P^{\rm th}=\sum_{m=-j}^j\frac{e^{-m\beta}}{Z_\beta}|j,m\rangle_{\rm opt}\langle j,m|,
\end{equation}
where $Z_\beta={\rm Tr}[\exp(-\beta J_{\rm opt})]$ is the partition function and $\beta\equiv\omega_P/(k_BT)$ is the dimensionless inverse temperature. Using Eqs.~(\ref{U}), the final unnormalized composite state can be written as
\begin{equation}
\rho(\theta)=U_{\theta,+}\rho_P^{\rm th}\otimes\rho_AU_{\theta,+}^\dagger
+U_{\theta,-}\rho_P^{\rm th}\otimes\rho_AU_{\theta,-}^\dagger,
\end{equation}
where the first and second terms correspond to the measurement results $|+\rangle$ and $|-\rangle$ about the ancillary qubit with probabilities $\mathcal{N}_\pm={\rm Tr}[U_{\theta,\pm}\rho_P^{\rm th}\otimes\rho_AU_{\theta,\pm}^\dagger]$, respectively, and $\rho_A$ is the initial state of ancillary qubit.

In the low-temperature regime, the population of the probe system is mostly on the ground state $|j,-j\rangle_{\rm opt}$. According to Eq.~(\ref{probability measurement result}), $\mathcal{N}_+\approx\mathcal{N}_{\theta,+}\rightarrow1$ in the large-$N$ limit. Thus, we can only consider the measurement outcome $|+\rangle$, i.e., $\rho(\theta)\approx U_{\theta,+}\rho_P^{\rm th}\otimes\rho_AU_{\theta,+}^\dagger$. Using the optimized conditions in Eqs.~(\ref{optimal g gz}), (\ref{optimal time}), (\ref{probeOptimal}), and (\ref{optimalFrequency}) and $\rho_A=|+\rangle\langle+|$, the normalized composite state can be approximately derived as
\begin{equation}
\begin{aligned}
\rho(\theta)&\approx\mathcal{N}_+^{-1}U_{\theta,+}\rho_P^{\rm th}\otimes\rho_AU_{\theta,+}^\dagger\\
&\approx\sum_{m=-j}^j\left(\frac{N-2m}{4N}\right)^2\frac{e^{-m\beta}}{Z_\beta\mathcal{N}_+}U_\theta^\prime|j,m\rangle_{\rm opt}\langle j,m|U_\theta^{\prime\dagger}\\
&\otimes|+\rangle\langle+|,
\end{aligned}
\end{equation}
where we have omitted the contribution from the raising and lowering operators $J_{\pm,x}$ in Eq.~(\ref{output state}) and then the effective evolution operator turns out to be
\begin{equation}
\begin{aligned}
U_\theta^\prime=&\frac{1-i(-1)^{n_P}}{2}U(t_2)R_x(\theta)\\
&\times\left[1+i(-1)^{n_1N+n_P}e^{-i\phi}e^{-i\pi J_z}\right].
\end{aligned}
\end{equation}
To have a unitary operator $U_\theta^\prime$, it is required that
\begin{equation}\label{unitary condition}
\begin{aligned}
&U_\theta^{\prime\dagger}U_\theta^\prime=\mathcal{I}^{N+1}+\frac{i(-1)^{n_1(N+1)+n_P}}{2} \\ \times&\left(e^{-i\phi}e^{-i\pi J_z}-e^{i\phi}e^{i\pi J_z}\right)=\mathcal{I}^{N+1},
\end{aligned}
\end{equation}
which is equivalent to specialize the longitudinal coupling strength:
\begin{equation}\label{longitudinal coupling}
g_z=\frac{N+2n_z-1}{2n_1+1}\frac{2\Delta_A}{(N+1)^2}
\end{equation}
with $n_z$ integer. Together with the optimized conditions in Eq.~(\ref{optimal g gz}), QFI with the thermal probe can be written in a compact form
\begin{equation}\label{QFI ana thermal}
\begin{aligned}
F_Q&=\sum_{m=-j}^j4p_m\langle j,m|J_x^2|j,m\rangle_{\rm opt}\\
&-\sum_{m,m'=-j}^j\frac{8p_mp_{m'}}{p_m+p_{m'}}|\langle j,m|e^{-i\pi J_z}J_x|j,m'\rangle_{\rm opt}|^2\\
&=\frac{4}{\mathcal{N}_+}\sum_{m=-j}^jm^2\left(\frac{N-2m}{4N}\right)^2
\frac{e^{-m\beta}}{Z_\beta}\\
&-\frac{8}{Z_\beta\mathcal{N}_+}\sum_{m=-j}^j\left( \frac{N-2m}{4N}\right)^2\frac{m^2}{e^{-m\beta}+e^{m\beta}},
\end{aligned}
\end{equation}
where $p_m=e^{-m\beta}(N-2m)^2/(16Z_\beta\mathcal{N}_+N^2)$.

\begin{figure}[htbp]
\begin{centering}
\includegraphics[width=0.8\linewidth]{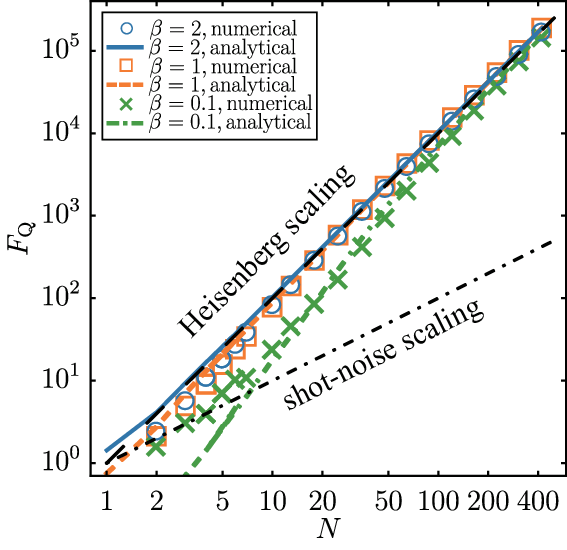}
\caption{QFI as a function of $N$ for a thermal state $\rho_P^{\rm th}$ with various inverse temperatures. The black-dashed line and the black dot-dashed line indicate the Heisenberg and shot-noise scalings, respectively. Here $\rho_A=|\varphi\rangle\langle\varphi|=|+\rangle\langle+|$, $t_1=(N+1)\pi/(4\Delta_A)$, $\omega_P/\Delta_A=40/(N+1)$, $g_z/\Delta_A\approx6/N$, and $g/\Delta_A\approx4\sqrt{2}/N$.}\label{QFI thermal state}
\end{centering}
\end{figure}

The analytical result in Eq.~(\ref{QFI ana thermal}), which holds the asymptotic square scaling law of $N$ under the large-$N$ and low-temperature limits, can be justified by the numerical simulation in Fig.~\ref{QFI thermal state} about QFI as a function of system size $N$ under various inverse temperatures. Without loss of generality, we choose $n_1=0$ in Eq.~(\ref{optimal time}), $n_P=10$ in Eq.~(\ref{optimalFrequency}), and $n_z=N$ in Eq.~(\ref{longitudinal coupling}), which yield $t_1=(N+1)\pi/(4\Delta_A)$, $\omega_P/\Delta_A=40/(N+1)$, and $g_z/\Delta_A=2(3N-1)/(N+1)^2\approx6/N$, respectively. Note here the dependence of $\omega_P$ and $g_z$ on $N$ does not mean they will vanish when $N\rightarrow\infty$. These values are merely used in numerical simulation over a finite size of probe. Subsequently, we have $g/\Delta_A=4\sqrt{2N(N-1)}/(N+1)^2\approx4\sqrt{2}/N$ according to Eqs.~(\ref{optimal g gz}). It is shown that a larger $\beta$ yields a behavior closer to the Heisenberg scaling. The numerical results become indistinguishable from the approximate analytical results for $N\geq10$, $N\geq15$, and $N\geq70$, when $\beta=2$, $\beta=1$, and $\beta=0.1$, respectively. Even in case of $\beta=0.1$, i.e., a high-temperature probe state, the scaling behavior of QFI approaches the Heisenberg limit for $N>200$.

\section{Effects of imprecise control}\label{imprecisecontrol}

It is generally difficult to control both encoding process associated with the to-be-estimated phase and to-be-optimized protocol parameters with an infinite accuracy. Our protocol is confirmed to be robust against the small deviation of the direction of the phase encoding in Eq.~(\ref{U}), the imperfection in the polarized direction for the probe state preparation in Eq.~(\ref{probeOptimal}), and the coupling strengths of both transversal and longitudinal interactions in Eq.~(\ref{optimal g gz}).

In Eq.~(\ref{U}), we have set $\alpha=x$ for the direction of phase encoding. With a small angle deviation from the $x$ axis, denoted by $\delta$, the encoding operator becomes
\begin{equation}
R'_x(\theta)=e^{-i\theta(J_x\cos\delta+J_y\sin\delta)}\approx e^{-i\theta[(1-\delta^2/2)J_x+\delta J_y]}
\end{equation}
up to the second order of $\delta$. With the ancillary qubit $|\varphi\rangle=|+\rangle$ and the optimal conditions of the probe state in Eq.~(\ref{probeOptimal}), and the joint-evolution time in Eq.~(\ref{optimal time}), the unnormalized state in Eq.~(\ref{output state}) deviates by replacing $R_x(\theta)$ with $R'_x(\theta)$. The relevant probability $\mathcal{N}_{\theta,\pm}$ is still given by Eq.~(\ref{probability measurement result}). Using Eqs.~(\ref{QFI pm}) and (\ref{QFI optimal}), the full quantum Fisher information under the large-$N$ limit reads,
\begin{equation}\label{QFIapp}
\begin{aligned}
F_Q&\approx\frac{N^2}{(N+1)^2}[(N+1)^2-4-\delta(-1)^{n_P}(2N-1)\\
&-\delta^2(N^2+2N-2)]
\end{aligned}
\end{equation}
up to the second order of $\delta$. Determined by $dF_Q/d\delta=0$, the maximum value of QFI is obtained when $\delta=(-1)^{n_P+1}(2N-1)/2(N^2+2N-2)$. The parity of the integer $n_P$ implies the optimal point of the deviation $\delta$ as well as the sign of the optimal $\delta$.

\begin{figure}[htbp]
\begin{centering}
\includegraphics[width=0.9\linewidth]{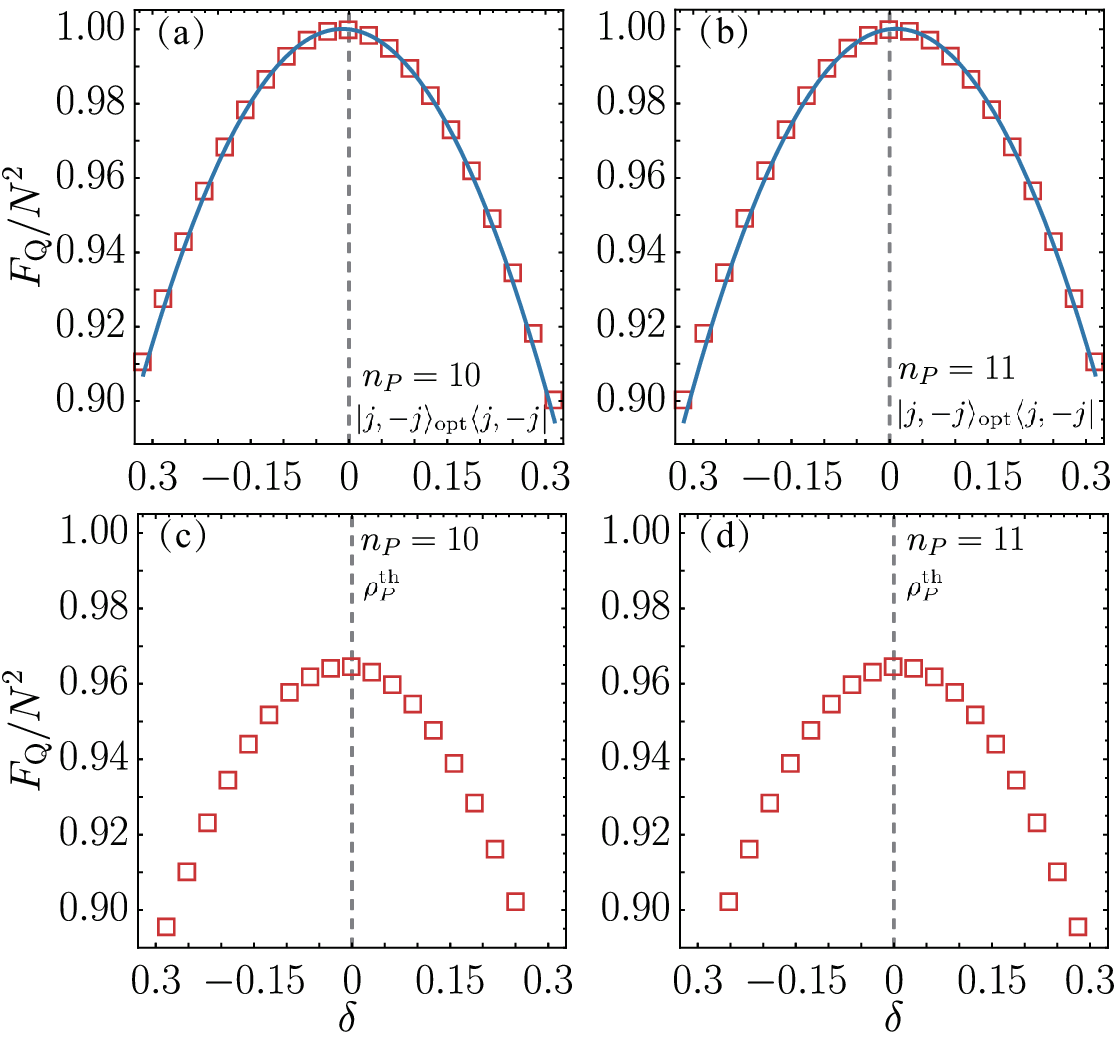}
\caption{Renormalized QFI $F_Q/N^2$ as a function of the deviation $\delta$ about encoding direction for the probe system prepared as (a) and (b) a polarized state $|\psi\rangle=|j,-j\rangle_{\rm opt}$ or (c) and (d) a thermal state $\rho^{\rm th}_P$ with $\beta=1$. The solid lines in (a) and (b) indicate the approximated analytical result in Eq.~(\ref{QFIapp}). The probe frequency $\omega_P$ is optimized by Eq.~(\ref{optimalFrequency}) with (a) and (c) an even number $n_P=10$ and (b) and (d) an odd number $n_P=11$. The probe size is $N=100$. $t_1=(N+1)\pi/(4\Delta_A)$, $g_z/\Delta_A\approx6/N$, and $g/\Delta_A\approx4\sqrt{2}/N$. }\label{QFI shift}
\end{centering}
\end{figure}

Figures~\ref{QFI shift}(a) and (b) show the dependence of QFI on the angle deviation $\delta$ under different parity of $n_P$ for a probe ensemble with a moderate size of $N=100$ prepared as $|\psi\rangle=|j,-j\rangle_{\rm opt}$. Due to Eq.~(\ref{optimalFrequency}), they have different $\omega_P$ that is determined by $n_P$. The other protocol parameters, including $t_1$, $g_z$, and $g$, are chosen the same as Fig.~\ref{QFI thermal state}. As expected from Eq.~(\ref{QFIapp}), QFI takes the peak value nearly $N^2$ around $\delta=0$ and exhibits an approximately symmetric distribution with a small right (left) shift for odd (even) $n_P$. For either $n_P$, QFI is insensitive to the angle deviation. $F_Q/N^2$ is found to be larger than $0.988$ even when $|\delta|=0.1$. Such an asymptotic Heisenberg-scaling behavior of QFI and its insensitivity to the imperfect encoding direction hold even if the probe starts from a mixed state. In Figs.~\ref{QFI shift}(c) and (d), we choose the thermal state in Eq.~(\ref{thermal state}) with $\beta=1$. It is found that $F_Q/N^2\approx0.964$ at $\delta\approx\mp0.01$ for even and odd $n_P$, respectively. And when $|\delta|=0.1$, it is still about $F_Q/N^2\approx0.956$.

In a practical scenario, the deviation varies with every run of the metrology protocol. We assume that $\delta$ follows a Gaussian distribution, e.g., $P(\delta)=(2\pi\sigma^2)^{-1/2}\exp[-\delta^2/(2\sigma^2)]$, where $\sigma$ denotes the standard deviation. Using the single-run QFI in Eq.~(\ref{QFIapp}), the averaged quantum Fisher information under the large-$N$ limit reads $F_{Q,{\rm ave}}=\int_{-\infty}^\infty d\delta P(\delta)F_Q=N^2/(N+1)^2[2\sigma^2-3+N(N+2)(1-\sigma^2)]$. When $\sigma=0$, it is exactly the same as Eq.~(\ref{QFIapp}) with $\delta=0$.

Suppose the initial probe state deviates from the optimized state $|j,-j\rangle_{\rm opt}$ by a small rotation about $J_z$, then the initial probe state can be written as $|\psi'\rangle=e^{i\varepsilon J_z}|j,-j\rangle_{\rm opt}$, where $\varepsilon$ is the deviation magnitude. Upon the ancillary qubit $|\varphi\rangle=|+\rangle$ and the optimal conditions in Eqs.~(\ref{optimal g gz}), (\ref{optimal time}), and (\ref{optimalFrequency}), the unnormalized state after measurement on the qubit and the relevant probability are similar to Eqs.~(\ref{output state}) and (\ref{probability measurement result}), respectively. Then using Eqs.~(\ref{QFI pm}) and (\ref{QFI optimal}), we have the full QFI under the large-$N$ limit up to the second order of $\varepsilon$:
\begin{equation}
F_Q\approx\frac{N^4}{(N+1)^2}\left[\frac{N+3}{N+1}-\varepsilon(-1)^{n_P}\frac{5}{N+1}-\varepsilon^2\frac{4N}{2N+3}\right].
\end{equation}
One can find that the maximum value of the QFI is obtained at $\varepsilon=(-1)^{n_P+1}5(2N+3)/8N(N+1)$. Similar to the result of the deviation $\delta$ in Eq.~(\ref{QFIapp}), the sign of the optimal $\varepsilon$ is determined by the parity of $n_P$ in Eq.~(\ref{optimalFrequency}). We always have $F_Q\approx N^2(1-2\varepsilon^2)$ when $N\rightarrow\infty$, confirming that our protocol is robust to the imperfect preparation.

\begin{figure}[htbp]
\begin{centering}
\includegraphics[width=0.8\linewidth]{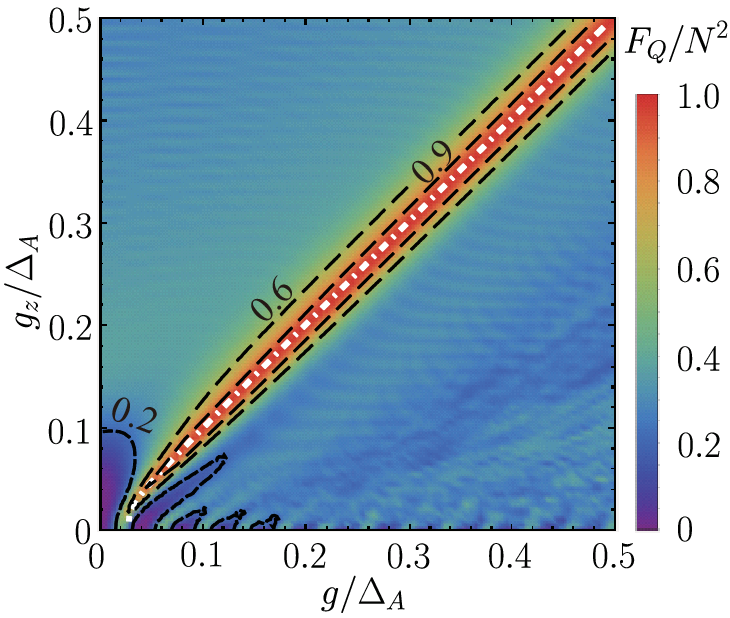}
\caption{Renormalized QFI $F_Q/N^2$ in the parametric space of $g/\Delta_A$ and $g_z/\Delta_A$ for the probe system prepared as the polarized state $|\psi\rangle=|j,-j\rangle_{\rm opt}$ with the probe spin number $N=100$. The white dot-dashed line presents the analytical result under the condition in Eq.~(\ref{optimal g gz}). Here $t_1=t_{1, {\rm opt}}(n_1=0)=(N+1)\pi/(4\Delta_A)$.}\label{QFI coupling strength}
\end{centering}
\end{figure}

All the preceding calculations of QFI are performed under the optimized condition about both transversal and longitudinal coupling strengths as given by Eq.~(\ref{optimal g gz}). In Fig.~\ref{QFI coupling strength}, we simulate the renormalized QFI $F_Q/N^2$ in a wide regime of both $g$ and $g_z$ to show the impact from the imprecise control over the protocol parameters. Here the probe system is initialized as the polarized state $|\psi\rangle=|j,-j\rangle_{\rm opt}$ with $N=100$; The regimes of coupling strengths are not relevant to $N$; and the joint-evolution duration of the first stage is optimized as $t_1=t_{1, {\rm opt}}(n_1=0)=(N+1)\pi/(4\Delta_A)$ as before. The white dot-dashed line indicates QFI of the maximum value $N^2$ obtained by the optimized coupling strengths satisfying Eq.~(\ref{optimal g gz}). The black dashed lines are the contour lines, by which one can find that the asymptotic quadratic scaling behavior of QFI remains valid when the probe-ancilla coupling strengths deviate from their optimal values with a certain magnitude. In particular, when the transversal coupling strength is fixed, e.g., $g/\Delta_A=0.2$ and $0.4$, we have $F_Q/N^2\geq0.9$ in the regimes $0.19\leq g_z/\Delta_A\leq0.21$ and $0.39\leq g_z/\Delta_A\leq0.41$, respectively. When the transversal and longitudinal coupling strengths are balanced, e.g., $g_z/\Delta_A=g/\Delta_A\approx0.043$ and $0.24$, we have $F_Q/N^2\approx0.84$ and $0.98$, respectively.

\section{Phase sensitivity based on parity detection}\label{Phasesensitivity}

From an operational perspective, it is important to quantify the ultimate achievable precision in parametric estimation by the measurement outcomes. The phase sensitivity~\cite{demkowicz2015quantum,bradshaw2018ultimate} characterizes the minimum statistical uncertainty about the estimated parameter under a given metrology protocol and hence provides a direct benchmark for comparing different metrology strategies. Theoretically, the phase sensitivity is lower bounded by the quantum Cram\'{e}r-Rao bound~\cite{helstrom1969quantum} which scales inversely with the square root of QFI. When the bound is saturated, all information about the to-be-estimated parameter encoded in the probe system has been extracted via the measurement outcomes. In addition, different measurements or detection methods applied to the same output result in different outcomes and hence phase sensitivities. Popular detection methods include homodyne detection~\cite{yuen1983noise,manceau2017improving}, intensity detection~\cite{szigeti2017pumped,ataman2018phase}, and parity detection~\cite{bollinger1996optimal,anisimov2010quantum,kielinski2024ghz,deng2024quantum}. In comparison to the others, parity detection reduces the measurement to a binary observable, i.e., even or odd parity, thereby enables more efficient extraction of phase information. Experimentally, parity detection can be realized in both atomic and photonic systems, through measuring the population in atomic ensembles~\cite{kielinski2024ghz} and the photon number of the cavity~\cite{deng2024quantum}, respectively. In this section, we derive the phase sensitivity of our protocol under the parity detection.

According to the error propagation equation~\cite{yurke19862}, the phase sensitivity under parity detection can be expressed as:
\begin{equation}\label{phase sensitivity}
|\Delta\theta|=\frac{\sqrt{1-\langle\Pi\rangle^2}}{\left|\partial \langle\Pi\rangle/\partial\theta\right|},
\end{equation}
where $\langle\Pi\rangle={\rm Tr}[\rho\Pi]$ denotes the output signal on the parity operator $\Pi$. The following derivation is performed under the optimized settings, i.e., the optimized joint-evolution time $t_{1,{\rm opt}}$, the probe eigenfrequency $\omega_P$, and the coupling strengths of the transversal and longitudinal interactions are set by Eq.~(\ref{optimal time}) with $n_1=0$, Eq.~(\ref{optimalFrequency}) with $n_P=10$, and Eq.~(\ref{optimal g gz}), respectively.

\subsection{Parity detection on the ancillary qubit}\label{detection on the qubit}

We first consider the parity detection performed on the ancillary qubit with the parity operator $\Pi=\Pi_A=\sigma_z$. In the circuit model~(\ref{protocol}), after the free joint evolution for Evolution 1, the parametric encoding, and the unconditional measurement, the composite system is in a separable state, with all information about the to-be-estimated parameter encoded solely in the probe system. It indicates that at this moment the parity detection on the ancillary qubit cannot extract any information. Therefore, it is necessary to incorporate the joint evolution for Evolution 2 to transfer the encoded information from the probe system to the ancillary system. It has been shown~\cite{pezze2017optimal,yurke19862,chen2024qubit} that such a transfer can be optimized when the joint evolution operator $U(t_2)$ becomes the time reversal of $U(t_{1, {\rm opt}})$, i.e.,
\begin{equation}\label{UTimeReversal}
U(t_2)=U^\dagger(t_{1,{\rm opt}})
\end{equation}
up to a global phase. With the optimized duration $t_1$ in Eq.~(\ref{optimal time}), Eq.~(\ref{UTimeReversal}) is equivalent to $U(t_{1,{\rm opt}})U(t_2)=e^{-iH(t_{1,{\rm opt}}+t_2)}=\mathcal{I}^{2(N+1)}$. It gives rise to
\begin{equation}\label{evolution time parity qubit}
t_2=(4n_2-1)t_{1,{\rm opt}}
\end{equation}
with $n_2$ integer and exactly the same longitudinal coupling strength $g_z$ in Eq.~(\ref{longitudinal coupling}) by using Eqs.~(\ref{optimal g gz}) and (\ref{optimalFrequency}).

With Eqs.~(\ref{output state}), (\ref{longitudinal coupling}), and (\ref{evolution time parity qubit}), one can obtain the output state of the composite system after two stages of evolution $U(t_2)|\Psi_{\theta,\pm}\rangle$ and hence,
\begin{equation}
\begin{aligned}
\langle\Pi_A\rangle=&\langle\Psi_{\theta,+}|U^\dagger(t_2)\sigma_zU(t_2)|\Psi_{\theta,+}\rangle\\
&+\langle\Psi_{\theta,-}|U^\dagger(t_2)\sigma_zU(t_2)|\Psi_{\theta,-}\rangle\\
=&\frac{N[N^2\cos(N\theta)+\cos\theta]}{(N+1)^2}.
\end{aligned}
\end{equation}
Then the phase sensitivity in Eq.~(\ref{phase sensitivity}) is obtained as
\begin{equation}\label{sensitivity qubit}
|\Delta\theta|=\frac{1}{N}\frac{\sqrt{(1+N^{-1})^4-[\sin(N\theta)+N^{-1}\sin\theta]^2}}{|\cos(N\theta)+N^{-2}\cos\theta|}.
\end{equation}
Its lower bound can be obtained by $d|\Delta\theta|/d\theta=0$, which yields the optimal working points for the estimated parameter:
\begin{equation}\label{optimal to-be-estimated phase qubit}
\theta_{\rm opt, A}=k\pi
\end{equation}
with $k$ integer. The relevant phase sensitivity is
\begin{equation}\label{sensitivity qubit minimum}
|\Delta\theta|_{\rm min, A}=\frac{(N+1)^2}{N(N^2+1)}.
\end{equation}
For a large-size probe with $N\gg1$, the phase sensitivity in Eq.~(\ref{sensitivity qubit}) is approximated as $|\Delta\theta|\approx1/N$ when $\cos\theta\neq0$, indicating HL. In addition, by substituting $\theta=\theta_{\rm opt, A}+\delta\theta$ with $|\delta\theta/\theta_{\rm opt, A}|\ll1$, the phase sensitivity around the optimal working points is
\begin{equation}
|\Delta\theta|\approx\frac{(N+1)^2}{N(N^2+1)}+\frac{4N+3}{2N+4}\left(\delta\theta\right)^2
\end{equation}
up to the second order in the phase deviation $\delta\theta$. One can find that the first order in the phase deviation $\delta\theta$ vanishes, confirming that our result about the to-be-estimated parameter $\theta_{\rm opt, A}$ in Eq.~(\ref{optimal to-be-estimated phase qubit}) is both optimal and insensitive to the deviation of $\theta$.

\begin{figure}[htbp]
\begin{centering}
\includegraphics[width=0.7\linewidth]{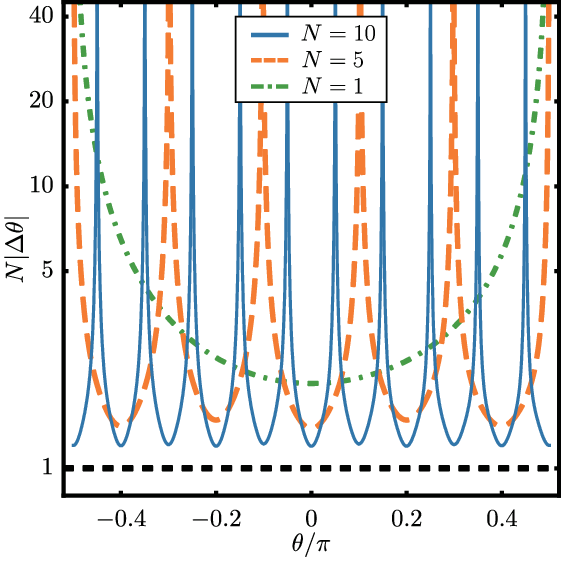}
\caption{Renormalized phase sensitivity $N|\Delta\theta|$ under the parity detection on the ancillary system. The black-dashed line represents HL. The probe is initialized as $|\psi\rangle=|j,-j\rangle_{\rm opt}$ and the evolution time of Stage 2 is $t_2=3t_{1,{\rm opt}}(n_1=0)$. The other parameters are the same as Fig.~\ref{QFI shift}.}\label{Renormalized phase sensitivity qubit}
\end{centering}
\end{figure}

This result can be demonstrated by numerical evaluations of the renormalized phase sensitivity $N|\Delta\theta|$ as a function of the to-be-estimated parameter $\theta$ in Fig.~\ref{Renormalized phase sensitivity qubit}, where the joint evolution time of the second stage $t_2$ is optimized by Eq.~(\ref{evolution time parity qubit}) with $n_2=1$. The horizontal line indicates the Heisenberg limit, which represents the ultimate precision achievable in phase estimation. One can find that the phase sensitivity exhibits an almost periodic behavior with the parameter $\theta$. The values of the phase sensitivity at the local-minimal points are nearly identical to the global-minimal point captured by Eq.~(\ref{optimal to-be-estimated phase qubit}) for a fixed probe size $N$. For example, when $N=5$, we have $N|\Delta\theta|\approx1.38$, $1.48$, and $1.41$ at $\theta/\pi=0$, $0.2$, and $0.4$, respectively. As the probe spin number $N$ increases, the optimal phase sensitivity at $\theta_{\rm opt}=0$ approaches the Heisenberg limit (see the horizontal line). In particular, we have $N|\Delta\theta|_{\rm min,A}\approx1.8$, $1.38$, and $1.20$ for $N=2$, $5$, and $10$, respectively, as expected by Eq.~(\ref{sensitivity qubit minimum}). When the parameter $\theta$ deviates from the optimal working point $\theta_{\rm opt}$, the phase sensitivity remains close to its optimal value, e.g., we have $|\theta|\leq0.110\pi$, $0.041\pi$, and $0.021\pi$ for $N=2$, $5$, and $10$, respectively, by holding $|\Delta\theta|\leq1.2|\Delta\theta|_{\rm min,A}$.

\subsection{Parity detection on the probe system}\label{detection on the probe}

If the parity detection is performed on the probe system, then the parity operator becomes $\Pi=\Pi_P=(-1)^{j-J_z}$, by which the extractable information varies with the joint-evolution time $t_2$.

In the case of $t_2=0$, the output state is the same as Eq.~(\ref{output state}), which indicates that the two subsystems are uncorrelated and the probe system is at a GHZ-type state. Consequently, one can obtain the output signal
\begin{equation}\label{signal probe zero}
\langle\Pi_P\rangle=\frac{N}{N+1}\sin(N\theta+j\pi+\phi),
\end{equation}
where the parameter $\phi$ is given by Eq.~(\ref{parameter phi}). Substituting Eq.~(\ref{signal probe zero}) to Eq.~(\ref{phase sensitivity}), we have
\begin{equation}\label{sensitivity probe zero}
|\Delta\theta|=\frac{1}{N}\sqrt{1+\frac{2N+1}{N^2\cos^2(N\theta+j\pi+\phi)}}.
\end{equation}
Similar to Sec.~\ref{detection on the qubit}, one can obtain the minimum value of the phase sensitivity
\begin{equation}\label{sensitivity probe minimum}
|\Delta\theta|_{\rm min,P}=\frac{N+1}{N^2},
\end{equation}
and the corresponding optimal working points
\begin{equation}\label{optimal to-be-estimated phase probe}
\theta_{\rm opt, P}=\frac{1}{N}[(k_1-j)\pi-\phi]
\end{equation}
with $k_1$ integer. It is interesting to find that the optimal working points $\theta_{\rm opt, P}$ in Eq.~(\ref{optimal to-be-estimated phase probe}) depend on the parameter $\phi$, or more precisely, on the ratio of the coupling strength and the detuning $g_z/\Delta_A$ as indicated by Eq.~(\ref{parameter phi}). It implies that, by only setting different ratio $g_z/\Delta_A$, one can shift the optimal working points $\theta_{\rm opt, P}$ and thereby achieve an optimal metrology over the full range of the to-be-estimated phase. In addition, by substituting $\theta=\theta_{\rm opt, P}+\delta\theta$ with $|\delta\theta/\theta_{\rm opt, P}|\ll1$, the phase sensitivity in Eq.~(\ref{sensitivity probe zero}) can be written as
\begin{equation}
|\Delta\theta|\approx\frac{N+1}{N^2}+\frac{2N+1}{2N+2}\left(\delta\theta\right)^2
\end{equation}
up to the second order of the phase deviation $\delta\theta$. It confirms that our result about the to-be-estimated parameter $\theta_{\rm opt, P}$ in Eq.~(\ref{optimal to-be-estimated phase probe}) is optimal.

In the case of $t_2\neq0$, the probe system becomes entangled again with the ancillary qubit, causing the information about the estimated parameter to be distributed across the two subsystems. An effective approach to enhance the ultimate precision is to drive the composite system back to a separable state, i.e., $U(t_2)=\mathcal{I}^{2(N+1)}$. Consequently, one can find the longitudinal coupling strength is that given by Eq.~(\ref{longitudinal coupling}) and the exact solution of the evolution time is
\begin{equation}\label{evolution time parity probe}
t_2=4n_3t_{1,{\rm opt}}
\end{equation}
with $n_3$ nonzero integer, where the optimized joint evolution time $t_{1, {\rm opt}}$ is given by Eq.~(\ref{optimal time}). Using Eqs.~(\ref{output state}), (\ref{longitudinal coupling}), and (\ref{evolution time parity probe}), the output signal becomes
\begin{equation}\label{signal probe nonzero}
\langle\Pi_P\rangle=(-1)^{N+n_z}\frac{N}{N+1}\sin(N\theta).
\end{equation}
Substituting Eq.~(\ref{signal probe nonzero}) to Eq.~(\ref{phase sensitivity}), we have the phase sensitivity
\begin{equation}\label{sensitivity probe nonzero}
|\Delta\theta|=\frac{1}{N}\sqrt{1+\frac{2N+1}{N^2\cos^2(N\theta)}}.
\end{equation}
One can find that the result in Eq.~(\ref{sensitivity probe nonzero}) is exactly the same as Eq.~(\ref{sensitivity probe zero}) with $\phi=-j\pi$. Thus, one can locate the minimum value of the phase
sensitivity in Eq.~(\ref{sensitivity probe minimum}) and the optimal working points $\theta_{\rm opt, P}=k_1\pi/N$ according to Eq.~(\ref{optimal to-be-estimated phase probe}).

\begin{figure}[htbp]
\begin{centering}
\includegraphics[width=0.8\linewidth]{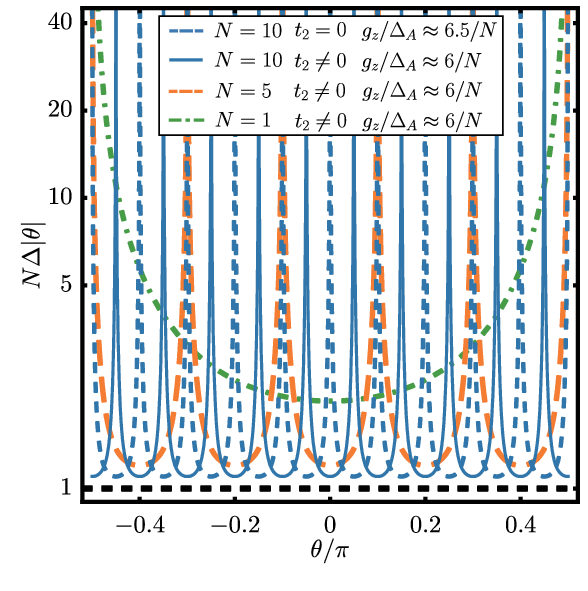}
\caption{Renormalized phase sensitivity $N|\Delta\theta|$ under the parity detection on the probe system with various $N$. The black-dashed line indicates the Heisenberg limit. The probe-spin ensemble is initialized as $|\psi\rangle=|j,-j\rangle_{\rm opt}$ and the evolution time of Stage 2 is set as $t_2=0$ or $t_2=4t_{1,{\rm opt}}(n_1=0)$. The other parameters are the same as Fig.~\ref{QFI shift}.}\label{Renormalized phase sensitivity probe}
\end{centering}
\end{figure}

The complementarity of $t_2=0$ and $t_2\neq0$ can be found in Fig.~\ref{Renormalized phase sensitivity probe}, that demonstrates the phase sensitivity about $\theta$ under the parity detection on the probe system. As indicated by the red dot-dashed line and blue-dashed line with the ratio $g_z/\Delta_A\approx6.5/N$ and $6.25/N$, respectively, the divergent regimes of the phase sensitivity can be avoided. Then the phase estimation over the full regime can be optimized provided that one can set two slightly different longitudinal coupling strength $g_z$. The green dot-dashed line, the yellow dashed line, and the black solid line are used to show the dependence of the renormalized phase sensitivity on the probe size $N$ with $t_2=4t_{1,{\rm opt}}(n_1=0)$. Different from Fig.~\ref{Renormalized phase sensitivity qubit}, the local local-minimal points of the phase sensitivity are also the global-minimal points as suggested by Eq.~(\ref{optimal to-be-estimated phase probe}). For example, when $N=5$, we have $N|\Delta\theta|=1.2$ at $\theta/\pi=0$, $\pm0.2$, and $\pm0.4$. As the probe size $N$ increases, the phase sensitivity at the optimal working points $\theta_{\rm opt, P}$ approaches HL. In particular, we have $N|\Delta\theta|_{\rm min,P}=1.5$, $1.2$, and $1.1$ for $N=2$, $5$, and $10$, respectively. In comparison to Fig.~\ref{Renormalized phase sensitivity qubit}, the last two results are lower than that obtained by the parity detection on the ancillary qubit. More important for the parity detection on the probe system, we find a wider regime about $\theta$ over which the phase sensitivity remains nearby its optimal value. To hold $|\Delta\theta|\leq1.2|\Delta\theta|_{\rm min,P}$, we have $|\theta|\leq0.117\pi$, $0.056\pi$, and $0.032\pi$ for $N=2$, $5$, and $10$, respectively.

\section{conclusion}\label{conclusion}

In summary, we incorporate the unconditional measurement on ancillary qubit to the phase estimation over a probe spin ensemble. The two components are weakly coupled to each other via a general Heisenberg XXZ interaction, that does not constrain the magnitude of ancillary-qubit frequency as in the previous work on ZZ interaction~\cite{ChenJing2025}. It is a metrology protocol without large-scale entanglement, nonlinear Hamiltonian, and strong or tunable external coupling strength. In our measurement-based metrology protocol, either polarized state or thermal state with a large probe size $N$ can be used to achieve an exact or asymptotic Heisenberg-scaling behavior $N^2$ in parameter estimation. We show that this quadratic scaling behavior is insensitive to the precise controls over the direction of the phase encoding, the probe state initialization, and the coupling strengths. To achieve the ultimate precision in a more efficient way, we propose to perform the parity detection on either probe or ancillary qubit. It turns out that by both methods, the phase sensitivity can approach the Heisenberg limit at the optimal working points. Under certain conditions, one can achieve optimal metrology precision over the entire regime of the to-be-estimated phase. In essence, our work paves an economical way toward the Heisenberg-scaling metrology.

\section*{Acknowledgments}

We acknowledge grant support from the National Natural Science Foundation of China (Grant No. U25A20199) and the ``Pioneer'' and ``Leading Goose'' R\&D of Zhejiang Province (Grant No. 2025C01028).

\appendix

\section{Time evolution operator for the XXZ Hamiltonian}\label{app}

This appendix contributes to deriving the unitary evolution operator for the full Hamiltonian in Eq.~(\ref{H}), which is recalled as
\begin{equation}\label{HXXZ}
\begin{aligned}
& H=H_0+H_I, \\
& H_0=\omega_PJ_z+\omega_A\sigma_z, \\
& H_I=g_zJ_z\sigma_z+g(J_x\sigma_x+J_y\sigma_y).
\end{aligned}
\end{equation}
Due to the commutation relation $[H, J_z+\sigma_z/2]=0$, the time evolution operator for $H$ can be partitioned into the subspaces $|j,j,e\rangle\oplus|j,-j,g\rangle\oplus\{|j,m+1,g\rangle,|j,m,e\rangle\}$ with $-j\leq m\leq j-1$ for quantum number $j=N/2$. $N$ is the size of the probe. The Hamiltonian in a typical subspace $\{|j,m+1,g\rangle,|j,m,e\rangle\}$ can be written as
\begin{equation}
H_m=\left(m+\frac{1}{2}\right)\omega_P-\frac{1}{2}g_z+\left(
\begin{array}{cc}
-\Lambda_m & \omega_m \\
\omega_m & \Lambda_m
\end{array}
\right),
\end{equation}
with $\omega_m=g\sqrt{(j-m)(j+m+1)}$, $\Lambda_m=(m+1/2)g_z+\Delta_A$, and $\Delta_A\equiv\omega_A-\omega_P/2$. The evolution operator in the relevant subspace thus reads
\begin{equation}
\begin{aligned}
&e^{-iH_mt}=e^{-i[(m+1/2)\omega_P-g_z/2]t}\times \\
&\left(
\begin{array}{cc}
\cos(\Omega_mt)+\frac{i\Lambda_m}{\Omega_m}\sin(\Omega_mt)&-\frac{i\omega_m}{\Omega_m}\sin(\Omega_mt)\\
-\frac{i\omega_m}{\Omega_m}\sin(\Omega_mt)&\cos(\Omega_mt)-\frac{i\Lambda_m}{\Omega_m}\sin(\Omega_mt)
\end{array}
\right),
\end{aligned}
\end{equation}
with $\Omega_m=\sqrt{\Lambda_m^2+\omega_m^2}$.

Consequently, the time evolution operator in the whole Hilbert space can be organized as
\begin{equation}\label{originalEvolution}
\begin{aligned}
e^{-iHt}&=\sum_{m=-j}^{j-1}e^{-iH_mt}+e^{-i[j(\omega_P+g_z)+\omega_A]t}|j,j,e\rangle\langle j,j,e|\\
&+e^{i[j(\omega_P-g_z)+\omega_A]t}|j,-j,g\rangle\langle j,-j,g|.
\end{aligned}
\end{equation}
Using the functions of the operator $J_z$, the full evolution operator can be rewritten in a more explicit way:
\begin{equation}
\begin{aligned}
e^{-iHt}&=\Bigg\{\left[\cos\Omega(J_z)t-\frac{i\Lambda(J_z)}{\Omega(J_z)}\sin\Omega(J_z)t\right]\otimes|e\rangle\langle e|\\
&-ie^{i\omega_Pt}g\frac{\sin\Omega(J_z)t}{\Omega(J_z)}J_-\otimes|e\rangle\langle g|\\
&-igJ_+\frac{\sin\Omega(J_z)t}{\Omega(J_z)}\otimes|g\rangle\langle e|\\
&+\left[\cos\Omega(J_z-1)t+\frac{i\Lambda(J_z-1)}{\Omega(J_z-1)}\sin\Omega(J_z-1)t\right]\\
&\otimes|g\rangle\langle g|e^{i\omega_Pt}\Bigg\}e^{-iA(J_z)t},
\end{aligned}
\end{equation}
where
\begin{subequations}
\begin{align}
\omega(J_z)&\equiv g\sqrt{(j-J_z)(j+J_z+1)},\\
\Lambda(J_z)&\equiv g_z\left( J_z+\frac{1}{2}\right)+\Delta_A, \\
\Omega(J_z)&\equiv \sqrt{\omega^2(J_z)+\Lambda^2(J_z)},\\
A(J_z)&\equiv \omega_P\left( J_z+\frac{1}{2}\right)-\frac{g_z}{2},
\end{align}
\end{subequations}
as given in Eqs.~(\ref{function omega}), (\ref{function lambda}), (\ref{function Omega}), and (\ref{function A}) in the main text.

\bibliographystyle{apsrevlong}
\bibliography{ref}

\begin{thebibliography}{66}%
\makeatletter
\providecommand \@ifxundefined [1]{%
 \@ifx{#1\undefined}
}%
\providecommand \@ifnum [1]{%
 \ifnum #1\expandafter \@firstoftwo
 \else \expandafter \@secondoftwo
 \fi
}%
\providecommand \@ifx [1]{%
 \ifx #1\expandafter \@firstoftwo
 \else \expandafter \@secondoftwo
 \fi
}%
\providecommand \natexlab [1]{#1}%
\providecommand \enquote  [1]{``#1''}%
\providecommand \bibnamefont  [1]{#1}%
\providecommand \bibfnamefont [1]{#1}%
\providecommand \citenamefont [1]{#1}%
\providecommand \href@noop [0]{\@secondoftwo}%
\providecommand \href [0]{\begingroup \@sanitize@url \@href}%
\providecommand \@href[1]{\@@startlink{#1}\@@href}%
\providecommand \@@href[1]{\endgroup#1\@@endlink}%
\providecommand \@sanitize@url [0]{\catcode `\\12\catcode `\$12\catcode
  `\&12\catcode `\#12\catcode `\^12\catcode `\_12\catcode `\%12\relax}%
\providecommand \@@startlink[1]{}%
\providecommand \@@endlink[0]{}%
\providecommand \url  [0]{\begingroup\@sanitize@url \@url }%
\providecommand \@url [1]{\endgroup\@href {#1}{\urlprefix }}%
\providecommand \urlprefix  [0]{URL }%
\providecommand \Eprint [0]{\href }%
\providecommand \doibase [0]{http://dx.doi.org/}%
\providecommand \selectlanguage [0]{\@gobble}%
\providecommand \bibinfo  [0]{\@secondoftwo}%
\providecommand \bibfield  [0]{\@secondoftwo}%
\providecommand \translation [1]{[#1]}%
\providecommand \BibitemOpen [0]{}%
\providecommand \bibitemStop [0]{}%
\providecommand \bibitemNoStop [0]{.\EOS\space}%
\providecommand \EOS [0]{\spacefactor3000\relax}%
\providecommand \BibitemShut  [1]{\csname bibitem#1\endcsname}%
\let\auto@bib@innerbib\@empty
\bibitem [{\citenamefont {Giovannetti}\ \emph {et~al.}(2004)\citenamefont
  {Giovannetti}, \citenamefont {Lloyd},\ and\ \citenamefont
  {Maccone}}]{giovannetti2004quantum}%
  \BibitemOpen
  \bibfield  {author} {\bibinfo {author} {\bibfnamefont {V.}~\bibnamefont
  {Giovannetti}}, \bibinfo {author} {\bibfnamefont {S.}~\bibnamefont {Lloyd}},
  \ and\ \bibinfo {author} {\bibfnamefont {L.}~\bibnamefont {Maccone}},\
  }\bibfield  {title} {\emph {\bibinfo {title} {Quantum-enhanced measurements:
  beating the standard quantum limit},\ }}\href {\doibase
  10.1126/science.1104149} {\bibfield  {journal} {\bibinfo  {journal}
  {Science}\ }\textbf {\bibinfo {volume} {306}},\ \bibinfo {pages} {1330}
  (\bibinfo {year} {2004})}\BibitemShut {NoStop}%
\bibitem [{\citenamefont {Sun}\ \emph {et~al.}(2010)\citenamefont {Sun},
  \citenamefont {Ma}, \citenamefont {Lu},\ and\ \citenamefont
  {Wang}}]{Sun2010fisher}%
  \BibitemOpen
  \bibfield  {author} {\bibinfo {author} {\bibfnamefont {Z.}~\bibnamefont
  {Sun}}, \bibinfo {author} {\bibfnamefont {J.}~\bibnamefont {Ma}}, \bibinfo
  {author} {\bibfnamefont {X.-M.}\ \bibnamefont {Lu}}, \ and\ \bibinfo {author}
  {\bibfnamefont {X.-G.}\ \bibnamefont {Wang}},\ }\bibfield  {title} {\emph
  {\bibinfo {title} {{Fisher} information in a quantum-critical environment},\
  }}\href {\doibase 10.1103/PhysRevA.82.022306} {\bibfield  {journal} {\bibinfo
   {journal} {Phys. Rev. A}\ }\textbf {\bibinfo {volume} {82}},\ \bibinfo
  {pages} {022306} (\bibinfo {year} {2010})}\BibitemShut {NoStop}%
\bibitem [{\citenamefont {Ma}\ \emph {et~al.}(2011)\citenamefont {Ma},
  \citenamefont {Huang}, \citenamefont {Wang},\ and\ \citenamefont
  {Sun}}]{ma2011quantum}%
  \BibitemOpen
  \bibfield  {author} {\bibinfo {author} {\bibfnamefont {J.}~\bibnamefont
  {Ma}}, \bibinfo {author} {\bibfnamefont {Y.-X.}\ \bibnamefont {Huang}},
  \bibinfo {author} {\bibfnamefont {X.-G.}\ \bibnamefont {Wang}}, \ and\
  \bibinfo {author} {\bibfnamefont {C.~P.}\ \bibnamefont {Sun}},\ }\bibfield
  {title} {\emph {\bibinfo {title} {Quantum {Fisher} information of the
  {Greenberger-Horne-Zeilinger} state in decoherence channels},\ }}\href
  {\doibase 10.1103/PhysRevA.84.022302} {\bibfield  {journal} {\bibinfo
  {journal} {Phys. Rev. A}\ }\textbf {\bibinfo {volume} {84}},\ \bibinfo
  {pages} {022302} (\bibinfo {year} {2011})}\BibitemShut {NoStop}%
\bibitem [{\citenamefont {Genoni}\ \emph {et~al.}(2012)\citenamefont {Genoni},
  \citenamefont {Olivares}, \citenamefont {Brivio}, \citenamefont {Cialdi},
  \citenamefont {Cipriani}, \citenamefont {Santamato}, \citenamefont
  {Vezzoli},\ and\ \citenamefont {Paris}}]{genoni2012optical}%
  \BibitemOpen
  \bibfield  {author} {\bibinfo {author} {\bibfnamefont {M.~G.}\ \bibnamefont
  {Genoni}}, \bibinfo {author} {\bibfnamefont {S.}~\bibnamefont {Olivares}},
  \bibinfo {author} {\bibfnamefont {D.}~\bibnamefont {Brivio}}, \bibinfo
  {author} {\bibfnamefont {S.}~\bibnamefont {Cialdi}}, \bibinfo {author}
  {\bibfnamefont {D.}~\bibnamefont {Cipriani}}, \bibinfo {author}
  {\bibfnamefont {A.}~\bibnamefont {Santamato}}, \bibinfo {author}
  {\bibfnamefont {S.}~\bibnamefont {Vezzoli}}, \ and\ \bibinfo {author}
  {\bibfnamefont {M.~G.~A.}\ \bibnamefont {Paris}},\ }\bibfield  {title} {\emph
  {\bibinfo {title} {Optical interferometry in the presence of large phase
  diffusion},\ }}\href {\doibase 10.1103/PhysRevA.85.043817} {\bibfield
  {journal} {\bibinfo  {journal} {Phys. Rev. A}\ }\textbf {\bibinfo {volume}
  {85}},\ \bibinfo {pages} {043817} (\bibinfo {year} {2012})}\BibitemShut
  {NoStop}%
\bibitem [{\citenamefont {Escher}\ \emph {et~al.}(2012)\citenamefont {Escher},
  \citenamefont {Davidovich}, \citenamefont {Zagury},\ and\ \citenamefont
  {de~Matos~Filho}}]{escher2012quantum}%
  \BibitemOpen
  \bibfield  {author} {\bibinfo {author} {\bibfnamefont {B.~M.}\ \bibnamefont
  {Escher}}, \bibinfo {author} {\bibfnamefont {L.}~\bibnamefont {Davidovich}},
  \bibinfo {author} {\bibfnamefont {N.}~\bibnamefont {Zagury}}, \ and\ \bibinfo
  {author} {\bibfnamefont {R.~L.}\ \bibnamefont {de~Matos~Filho}},\ }\bibfield
  {title} {\emph {\bibinfo {title} {Quantum metrological limits via a
  variational approach},\ }}\href {\doibase 10.1103/PhysRevLett.109.190404}
  {\bibfield  {journal} {\bibinfo  {journal} {Phys. Rev. Lett.}\ }\textbf
  {\bibinfo {volume} {109}},\ \bibinfo {pages} {190404} (\bibinfo {year}
  {2012})}\BibitemShut {NoStop}%
\bibitem [{\citenamefont {Zhong}\ \emph {et~al.}(2013)\citenamefont {Zhong},
  \citenamefont {Sun}, \citenamefont {Ma}, \citenamefont {Wang},\ and\
  \citenamefont {Nori}}]{zhong2013fisher}%
  \BibitemOpen
  \bibfield  {author} {\bibinfo {author} {\bibfnamefont {W.}~\bibnamefont
  {Zhong}}, \bibinfo {author} {\bibfnamefont {Z.}~\bibnamefont {Sun}}, \bibinfo
  {author} {\bibfnamefont {J.}~\bibnamefont {Ma}}, \bibinfo {author}
  {\bibfnamefont {X.}~\bibnamefont {Wang}}, \ and\ \bibinfo {author}
  {\bibfnamefont {F.}~\bibnamefont {Nori}},\ }\bibfield  {title} {\emph
  {\bibinfo {title} {{Fisher} information under decoherence in bloch
  representation},\ }}\href {\doibase 10.1103/PhysRevA.87.022337} {\bibfield
  {journal} {\bibinfo  {journal} {Phys. Rev. A}\ }\textbf {\bibinfo {volume}
  {87}},\ \bibinfo {pages} {022337} (\bibinfo {year} {2013})}\BibitemShut
  {NoStop}%
\bibitem [{\citenamefont {Katori}(2011)}]{Katori2011optical}%
  \BibitemOpen
  \bibfield  {author} {\bibinfo {author} {\bibfnamefont {H.}~\bibnamefont
  {Katori}},\ }\bibfield  {title} {\emph {\bibinfo {title} {Optical lattice
  clocks and quantum metrology},\ }}\href {\doibase
  https://doi.org/10.1038/nphoton.2011.45} {\bibfield  {journal} {\bibinfo
  {journal} {Nat. Photon.}\ }\textbf {\bibinfo {volume} {5}},\ \bibinfo {pages}
  {203} (\bibinfo {year} {2011})}\BibitemShut {NoStop}%
\bibitem [{\citenamefont {Ludlow}\ \emph {et~al.}(2015)\citenamefont {Ludlow},
  \citenamefont {Boyd}, \citenamefont {Ye}, \citenamefont {Peik},\ and\
  \citenamefont {Schmidt}}]{Ludlow2015optical}%
  \BibitemOpen
  \bibfield  {author} {\bibinfo {author} {\bibfnamefont {A.~D.}\ \bibnamefont
  {Ludlow}}, \bibinfo {author} {\bibfnamefont {M.~M.}\ \bibnamefont {Boyd}},
  \bibinfo {author} {\bibfnamefont {J.}~\bibnamefont {Ye}}, \bibinfo {author}
  {\bibfnamefont {E.}~\bibnamefont {Peik}}, \ and\ \bibinfo {author}
  {\bibfnamefont {P.~O.}\ \bibnamefont {Schmidt}},\ }\bibfield  {title} {\emph
  {\bibinfo {title} {Optical atomic clocks},\ }}\href {\doibase
  10.1103/RevModPhys.87.637} {\bibfield  {journal} {\bibinfo  {journal} {Rev.
  Mod. Phys.}\ }\textbf {\bibinfo {volume} {87}},\ \bibinfo {pages} {637}
  (\bibinfo {year} {2015})}\BibitemShut {NoStop}%
\bibitem [{\citenamefont {Caves}(1981)}]{Caves1981quantum}%
  \BibitemOpen
  \bibfield  {author} {\bibinfo {author} {\bibfnamefont {C.~M.}\ \bibnamefont
  {Caves}},\ }\bibfield  {title} {\emph {\bibinfo {title} {Quantum-mechanical
  noise in an interferometer},\ }}\href {\doibase 10.1103/PhysRevD.23.1693}
  {\bibfield  {journal} {\bibinfo  {journal} {Phys. Rev. D}\ }\textbf {\bibinfo
  {volume} {23}},\ \bibinfo {pages} {1693} (\bibinfo {year}
  {1981})}\BibitemShut {NoStop}%
\bibitem [{\citenamefont {Taylor}\ and\ \citenamefont
  {Bowen}(2016)}]{Taylor2016quantum}%
  \BibitemOpen
  \bibfield  {author} {\bibinfo {author} {\bibfnamefont {M.~A.}\ \bibnamefont
  {Taylor}}\ and\ \bibinfo {author} {\bibfnamefont {W.~P.}\ \bibnamefont
  {Bowen}},\ }\bibfield  {title} {\emph {\bibinfo {title} {Quantum metrology
  and its application in biology},\ }}\href {\doibase
  https://doi.org/10.1016/j.physrep.2015.12.002} {\bibfield  {journal}
  {\bibinfo  {journal} {Phys. Rep.}\ }\textbf {\bibinfo {volume} {615}},\
  \bibinfo {pages} {1} (\bibinfo {year} {2016})}\BibitemShut {NoStop}%
\bibitem [{\citenamefont {Mauranyapin}\ \emph {et~al.}(2017)\citenamefont
  {Mauranyapin}, \citenamefont {Madsen}, \citenamefont {Taylor}, \citenamefont
  {Waleed},\ and\ \citenamefont {Bowen}}]{Mauranyapin2017evanescent}%
  \BibitemOpen
  \bibfield  {author} {\bibinfo {author} {\bibfnamefont {N.}~\bibnamefont
  {Mauranyapin}}, \bibinfo {author} {\bibfnamefont {L.}~\bibnamefont {Madsen}},
  \bibinfo {author} {\bibfnamefont {M.}~\bibnamefont {Taylor}}, \bibinfo
  {author} {\bibfnamefont {M.}~\bibnamefont {Waleed}}, \ and\ \bibinfo {author}
  {\bibfnamefont {W.}~\bibnamefont {Bowen}},\ }\bibfield  {title} {\emph
  {\bibinfo {title} {Evanescent single-molecule biosensing with quantum-limited
  precision},\ }}\href {\doibase https://doi.org/10.1038/nphoton.2017.99}
  {\bibfield  {journal} {\bibinfo  {journal} {Nat. Photon.}\ }\textbf {\bibinfo
  {volume} {11}},\ \bibinfo {pages} {477} (\bibinfo {year} {2017})}\BibitemShut
  {NoStop}%
\bibitem [{\citenamefont {Jones}\ \emph {et~al.}(2009)\citenamefont {Jones},
  \citenamefont {Karlen}, \citenamefont {Fitzsimons}, \citenamefont {Ardavan},
  \citenamefont {Benjamin}, \citenamefont {Briggs},\ and\ \citenamefont
  {Morton}}]{Jones2009magnetic}%
  \BibitemOpen
  \bibfield  {author} {\bibinfo {author} {\bibfnamefont {J.~A.}\ \bibnamefont
  {Jones}}, \bibinfo {author} {\bibfnamefont {S.~D.}\ \bibnamefont {Karlen}},
  \bibinfo {author} {\bibfnamefont {J.}~\bibnamefont {Fitzsimons}}, \bibinfo
  {author} {\bibfnamefont {A.}~\bibnamefont {Ardavan}}, \bibinfo {author}
  {\bibfnamefont {S.~C.}\ \bibnamefont {Benjamin}}, \bibinfo {author}
  {\bibfnamefont {G.~A.~D.}\ \bibnamefont {Briggs}}, \ and\ \bibinfo {author}
  {\bibfnamefont {J.~J.}\ \bibnamefont {Morton}},\ }\bibfield  {title} {\emph
  {\bibinfo {title} {Magnetic field sensing beyond the standard quantum limit
  using 10-spin noon states},\ }}\href {\doibase
  https://www.science.org/doi/10.1126/science.1170730} {\bibfield  {journal}
  {\bibinfo  {journal} {Science}\ }\textbf {\bibinfo {volume} {324}},\ \bibinfo
  {pages} {1166} (\bibinfo {year} {2009})}\BibitemShut {NoStop}%
\bibitem [{\citenamefont {Kay}(1993)}]{kay1993fundamentals}%
  \BibitemOpen
  \bibfield  {author} {\bibinfo {author} {\bibfnamefont {S.~M.}\ \bibnamefont
  {Kay}},\ }\href@noop {} {\emph {\bibinfo {title} {Fundamentals of statistical
  signal processing: estimation theory}}}\ (\bibinfo  {publisher}
  {Prentice-Hall, Inc.},\ \bibinfo {year} {1993})\BibitemShut {NoStop}%
\bibitem [{\citenamefont {Giovannetti}\ \emph {et~al.}(2006)\citenamefont
  {Giovannetti}, \citenamefont {Lloyd},\ and\ \citenamefont
  {Maccone}}]{giovannetti2006quantum}%
  \BibitemOpen
  \bibfield  {author} {\bibinfo {author} {\bibfnamefont {V.}~\bibnamefont
  {Giovannetti}}, \bibinfo {author} {\bibfnamefont {S.}~\bibnamefont {Lloyd}},
  \ and\ \bibinfo {author} {\bibfnamefont {L.}~\bibnamefont {Maccone}},\
  }\bibfield  {title} {\emph {\bibinfo {title} {Quantum metrology},\ }}\href
  {\doibase 10.1103/PhysRevLett.96.010401} {\bibfield  {journal} {\bibinfo
  {journal} {Phys. Rev. Lett.}\ }\textbf {\bibinfo {volume} {96}},\ \bibinfo
  {pages} {010401} (\bibinfo {year} {2006})}\BibitemShut {NoStop}%
\bibitem [{\citenamefont {Song}\ \emph {et~al.}(2019)\citenamefont {Song},
  \citenamefont {Xu}, \citenamefont {Li}, \citenamefont {Zhang}, \citenamefont
  {Zhang}, \citenamefont {Liu}, \citenamefont {Guo}, \citenamefont {Wang},
  \citenamefont {Ren}, \citenamefont {Hao} \emph
  {et~al.}}]{song2019generation}%
  \BibitemOpen
  \bibfield  {author} {\bibinfo {author} {\bibfnamefont {C.}~\bibnamefont
  {Song}}, \bibinfo {author} {\bibfnamefont {K.}~\bibnamefont {Xu}}, \bibinfo
  {author} {\bibfnamefont {H.-K.}\ \bibnamefont {Li}}, \bibinfo {author}
  {\bibfnamefont {Y.-R.}\ \bibnamefont {Zhang}}, \bibinfo {author}
  {\bibfnamefont {X.}~\bibnamefont {Zhang}}, \bibinfo {author} {\bibfnamefont
  {W.-X.}\ \bibnamefont {Liu}}, \bibinfo {author} {\bibfnamefont {Q.-J.}\
  \bibnamefont {Guo}}, \bibinfo {author} {\bibfnamefont {Z.}~\bibnamefont
  {Wang}}, \bibinfo {author} {\bibfnamefont {W.-H.}\ \bibnamefont {Ren}},
  \bibinfo {author} {\bibfnamefont {J.}~\bibnamefont {Hao}},  \emph {et~al.},\
  }\bibfield  {title} {\emph {\bibinfo {title} {Generation of multicomponent
  atomic schr{\"o}dinger cat states of up to 20 qubits},\ }}\href {\doibase
  https://doi.org/10.1126/science.aay0600} {\bibfield  {journal} {\bibinfo
  {journal} {Science}\ }\textbf {\bibinfo {volume} {365}},\ \bibinfo {pages}
  {574} (\bibinfo {year} {2019})}\BibitemShut {NoStop}%
\bibitem [{\citenamefont {Chalopin}\ \emph {et~al.}(2018)\citenamefont
  {Chalopin}, \citenamefont {Bouazza}, \citenamefont {Evrard}, \citenamefont
  {Makhalov}, \citenamefont {Dreon}, \citenamefont {Dalibard}, \citenamefont
  {Sidorenkov},\ and\ \citenamefont {Nascimbene}}]{chalopin2018quantum}%
  \BibitemOpen
  \bibfield  {author} {\bibinfo {author} {\bibfnamefont {T.}~\bibnamefont
  {Chalopin}}, \bibinfo {author} {\bibfnamefont {C.}~\bibnamefont {Bouazza}},
  \bibinfo {author} {\bibfnamefont {A.}~\bibnamefont {Evrard}}, \bibinfo
  {author} {\bibfnamefont {V.}~\bibnamefont {Makhalov}}, \bibinfo {author}
  {\bibfnamefont {D.}~\bibnamefont {Dreon}}, \bibinfo {author} {\bibfnamefont
  {J.}~\bibnamefont {Dalibard}}, \bibinfo {author} {\bibfnamefont {L.~A.}\
  \bibnamefont {Sidorenkov}}, \ and\ \bibinfo {author} {\bibfnamefont
  {S.}~\bibnamefont {Nascimbene}},\ }\bibfield  {title} {\emph {\bibinfo
  {title} {Quantum-enhanced sensing using non-classical spin states of a highly
  magnetic atom},\ }}\href {\doibase 10.1038/s41467-018-07433-1} {\bibfield
  {journal} {\bibinfo  {journal} {Nat. Commun.}\ }\textbf {\bibinfo {volume}
  {9}},\ \bibinfo {pages} {4955} (\bibinfo {year} {2018})}\BibitemShut
  {NoStop}%
\bibitem [{\citenamefont {Kaufmann}\ \emph {et~al.}(2017)\citenamefont
  {Kaufmann}, \citenamefont {Ruster}, \citenamefont {Schmiegelow},
  \citenamefont {Luda}, \citenamefont {Kaushal}, \citenamefont {Schulz},
  \citenamefont {Von~Lindenfels}, \citenamefont {Schmidt-Kaler},\ and\
  \citenamefont {Poschinger}}]{kaufmann2017scalable}%
  \BibitemOpen
  \bibfield  {author} {\bibinfo {author} {\bibfnamefont {H.}~\bibnamefont
  {Kaufmann}}, \bibinfo {author} {\bibfnamefont {T.}~\bibnamefont {Ruster}},
  \bibinfo {author} {\bibfnamefont {C.~T.}\ \bibnamefont {Schmiegelow}},
  \bibinfo {author} {\bibfnamefont {M.~A.}\ \bibnamefont {Luda}}, \bibinfo
  {author} {\bibfnamefont {V.}~\bibnamefont {Kaushal}}, \bibinfo {author}
  {\bibfnamefont {J.}~\bibnamefont {Schulz}}, \bibinfo {author} {\bibfnamefont
  {D.}~\bibnamefont {Von~Lindenfels}}, \bibinfo {author} {\bibfnamefont
  {F.}~\bibnamefont {Schmidt-Kaler}}, \ and\ \bibinfo {author} {\bibfnamefont
  {U.}~\bibnamefont {Poschinger}},\ }\bibfield  {title} {\emph {\bibinfo
  {title} {Scalable creation of long-lived multipartite entanglement},\ }}\href
  {\doibase 10.1103/PhysRevLett.119.150503} {\bibfield  {journal} {\bibinfo
  {journal} {Phys. Rev. Lett.}\ }\textbf {\bibinfo {volume} {119}},\ \bibinfo
  {pages} {150503} (\bibinfo {year} {2017})}\BibitemShut {NoStop}%
\bibitem [{\citenamefont {Kitagawa}\ and\ \citenamefont
  {Ueda}(1993)}]{kitagawa1993squeezed}%
  \BibitemOpen
  \bibfield  {author} {\bibinfo {author} {\bibfnamefont {M.}~\bibnamefont
  {Kitagawa}}\ and\ \bibinfo {author} {\bibfnamefont {M.}~\bibnamefont
  {Ueda}},\ }\bibfield  {title} {\emph {\bibinfo {title} {Squeezed spin
  states},\ }}\href {\doibase 10.1103/PhysRevA.47.5138} {\bibfield  {journal}
  {\bibinfo  {journal} {Phys. Rev. A}\ }\textbf {\bibinfo {volume} {47}},\
  \bibinfo {pages} {5138} (\bibinfo {year} {1993})}\BibitemShut {NoStop}%
\bibitem [{\citenamefont {Wineland}\ \emph {et~al.}(1994)\citenamefont
  {Wineland}, \citenamefont {Bollinger}, \citenamefont {Itano},\ and\
  \citenamefont {Heinzen}}]{wineland1994squeezed}%
  \BibitemOpen
  \bibfield  {author} {\bibinfo {author} {\bibfnamefont {D.~J.}\ \bibnamefont
  {Wineland}}, \bibinfo {author} {\bibfnamefont {J.~J.}\ \bibnamefont
  {Bollinger}}, \bibinfo {author} {\bibfnamefont {W.~M.}\ \bibnamefont
  {Itano}}, \ and\ \bibinfo {author} {\bibfnamefont {D.~J.}\ \bibnamefont
  {Heinzen}},\ }\bibfield  {title} {\emph {\bibinfo {title} {Squeezed atomic
  states and projection noise in spectroscopy},\ }}\href {\doibase
  10.1103/PhysRevA.50.67} {\bibfield  {journal} {\bibinfo  {journal} {Phys.
  Rev. A}\ }\textbf {\bibinfo {volume} {50}},\ \bibinfo {pages} {67} (\bibinfo
  {year} {1994})}\BibitemShut {NoStop}%
\bibitem [{\citenamefont {Zhang}\ \emph {et~al.}(2013)\citenamefont {Zhang},
  \citenamefont {Li}, \citenamefont {Yang},\ and\ \citenamefont
  {Jin}}]{zhang2013quantum}%
  \BibitemOpen
  \bibfield  {author} {\bibinfo {author} {\bibfnamefont {Y.-M.}\ \bibnamefont
  {Zhang}}, \bibinfo {author} {\bibfnamefont {X.-W.}\ \bibnamefont {Li}},
  \bibinfo {author} {\bibfnamefont {W.}~\bibnamefont {Yang}}, \ and\ \bibinfo
  {author} {\bibfnamefont {G.-R.}\ \bibnamefont {Jin}},\ }\bibfield  {title}
  {\emph {\bibinfo {title} {Quantum {Fisher} information of entangled coherent
  states in the presence of photon loss},\ }}\href {\doibase
  10.1103/PhysRevA.88.043832} {\bibfield  {journal} {\bibinfo  {journal} {Phys.
  Rev. A}\ }\textbf {\bibinfo {volume} {88}},\ \bibinfo {pages} {043832}
  (\bibinfo {year} {2013})}\BibitemShut {NoStop}%
\bibitem [{\citenamefont {Wineland}\ \emph {et~al.}(1992)\citenamefont
  {Wineland}, \citenamefont {Bollinger}, \citenamefont {Itano}, \citenamefont
  {Moore},\ and\ \citenamefont {Heinzen}}]{wineland1992spin}%
  \BibitemOpen
  \bibfield  {author} {\bibinfo {author} {\bibfnamefont {D.~J.}\ \bibnamefont
  {Wineland}}, \bibinfo {author} {\bibfnamefont {J.~J.}\ \bibnamefont
  {Bollinger}}, \bibinfo {author} {\bibfnamefont {W.~M.}\ \bibnamefont
  {Itano}}, \bibinfo {author} {\bibfnamefont {F.~L.}\ \bibnamefont {Moore}}, \
  and\ \bibinfo {author} {\bibfnamefont {D.~J.}\ \bibnamefont {Heinzen}},\
  }\bibfield  {title} {\emph {\bibinfo {title} {Spin squeezing and reduced
  quantum noise in spectroscopy},\ }}\href {\doibase 10.1103/PhysRevA.46.R6797}
  {\bibfield  {journal} {\bibinfo  {journal} {Phys. Rev. A}\ }\textbf {\bibinfo
  {volume} {46}},\ \bibinfo {pages} {R6797} (\bibinfo {year}
  {1992})}\BibitemShut {NoStop}%
\bibitem [{\citenamefont {S{\o}rensen}\ \emph {et~al.}(2001)\citenamefont
  {S{\o}rensen}, \citenamefont {Duan}, \citenamefont {Cirac},\ and\
  \citenamefont {Zoller}}]{sorensen2001many}%
  \BibitemOpen
  \bibfield  {author} {\bibinfo {author} {\bibfnamefont {A.}~\bibnamefont
  {S{\o}rensen}}, \bibinfo {author} {\bibfnamefont {L.-M.}\ \bibnamefont
  {Duan}}, \bibinfo {author} {\bibfnamefont {J.~I.}\ \bibnamefont {Cirac}}, \
  and\ \bibinfo {author} {\bibfnamefont {P.}~\bibnamefont {Zoller}},\
  }\bibfield  {title} {\emph {\bibinfo {title} {Many-particle entanglement with
  bose--einstein condensates},\ }}\href {\doibase
  https://doi.org/10.1038/35051038} {\bibfield  {journal} {\bibinfo  {journal}
  {Nature}\ }\textbf {\bibinfo {volume} {409}},\ \bibinfo {pages} {63}
  (\bibinfo {year} {2001})}\BibitemShut {NoStop}%
\bibitem [{\citenamefont {Gross}\ \emph {et~al.}(2010)\citenamefont {Gross},
  \citenamefont {Zibold}, \citenamefont {Nicklas}, \citenamefont {Esteve},\
  and\ \citenamefont {Oberthaler}}]{gross2010nonlinear}%
  \BibitemOpen
  \bibfield  {author} {\bibinfo {author} {\bibfnamefont {C.}~\bibnamefont
  {Gross}}, \bibinfo {author} {\bibfnamefont {T.}~\bibnamefont {Zibold}},
  \bibinfo {author} {\bibfnamefont {E.}~\bibnamefont {Nicklas}}, \bibinfo
  {author} {\bibfnamefont {J.}~\bibnamefont {Esteve}}, \ and\ \bibinfo {author}
  {\bibfnamefont {M.~K.}\ \bibnamefont {Oberthaler}},\ }\bibfield  {title}
  {\emph {\bibinfo {title} {Nonlinear atom interferometer surpasses classical
  precision limit},\ }}\href {\doibase https://doi.org/10.1038/nature08919}
  {\bibfield  {journal} {\bibinfo  {journal} {Nature}\ }\textbf {\bibinfo
  {volume} {464}},\ \bibinfo {pages} {1165} (\bibinfo {year}
  {2010})}\BibitemShut {NoStop}%
\bibitem [{\citenamefont {Riedel}\ \emph {et~al.}(2010)\citenamefont {Riedel},
  \citenamefont {B{\"o}hi}, \citenamefont {Li}, \citenamefont {H{\"a}nsch},
  \citenamefont {Sinatra},\ and\ \citenamefont {Treutlein}}]{riedel2010atom}%
  \BibitemOpen
  \bibfield  {author} {\bibinfo {author} {\bibfnamefont {M.~F.}\ \bibnamefont
  {Riedel}}, \bibinfo {author} {\bibfnamefont {P.}~\bibnamefont {B{\"o}hi}},
  \bibinfo {author} {\bibfnamefont {Y.}~\bibnamefont {Li}}, \bibinfo {author}
  {\bibfnamefont {T.~W.}\ \bibnamefont {H{\"a}nsch}}, \bibinfo {author}
  {\bibfnamefont {A.}~\bibnamefont {Sinatra}}, \ and\ \bibinfo {author}
  {\bibfnamefont {P.}~\bibnamefont {Treutlein}},\ }\bibfield  {title} {\emph
  {\bibinfo {title} {Atom-chip-based generation of entanglement for quantum
  metrology},\ }}\href {\doibase https://doi.org/10.1038/nature08988}
  {\bibfield  {journal} {\bibinfo  {journal} {Nature}\ }\textbf {\bibinfo
  {volume} {464}},\ \bibinfo {pages} {1170} (\bibinfo {year}
  {2010})}\BibitemShut {NoStop}%
\bibitem [{\citenamefont {Schleier-Smith}\ \emph {et~al.}(2010)\citenamefont
  {Schleier-Smith}, \citenamefont {Leroux},\ and\ \citenamefont
  {Vuleti{\'c}}}]{schleier2010squeezing}%
  \BibitemOpen
  \bibfield  {author} {\bibinfo {author} {\bibfnamefont {M.~H.}\ \bibnamefont
  {Schleier-Smith}}, \bibinfo {author} {\bibfnamefont {I.~D.}\ \bibnamefont
  {Leroux}}, \ and\ \bibinfo {author} {\bibfnamefont {V.}~\bibnamefont
  {Vuleti{\'c}}},\ }\bibfield  {title} {\emph {\bibinfo {title} {Squeezing the
  collective spin of a dilute atomic ensemble by cavity feedback},\ }}\href
  {\doibase 10.1103/PhysRevA.81.021804} {\bibfield  {journal} {\bibinfo
  {journal} {Phys. Rev. A}\ }\textbf {\bibinfo {volume} {81}},\ \bibinfo
  {pages} {021804} (\bibinfo {year} {2010})}\BibitemShut {NoStop}%
\bibitem [{\citenamefont {Leroux}\ \emph {et~al.}(2010)\citenamefont {Leroux},
  \citenamefont {Schleier-Smith},\ and\ \citenamefont
  {Vuleti{\'c}}}]{leroux2010implementation}%
  \BibitemOpen
  \bibfield  {author} {\bibinfo {author} {\bibfnamefont {I.~D.}\ \bibnamefont
  {Leroux}}, \bibinfo {author} {\bibfnamefont {M.~H.}\ \bibnamefont
  {Schleier-Smith}}, \ and\ \bibinfo {author} {\bibfnamefont {V.}~\bibnamefont
  {Vuleti{\'c}}},\ }\bibfield  {title} {\emph {\bibinfo {title} {Implementation
  of cavity squeezing of a collective atomic spin},\ }}\href {\doibase
  10.1103/PhysRevLett.104.073602} {\bibfield  {journal} {\bibinfo  {journal}
  {Phys. Rev. Lett.}\ }\textbf {\bibinfo {volume} {104}},\ \bibinfo {pages}
  {073602} (\bibinfo {year} {2010})}\BibitemShut {NoStop}%
\bibitem [{\citenamefont {Norcia}\ \emph {et~al.}(2018)\citenamefont {Norcia},
  \citenamefont {Lewis-Swan}, \citenamefont {Cline}, \citenamefont {Zhu},
  \citenamefont {Rey},\ and\ \citenamefont {Thompson}}]{norcia2018cavity}%
  \BibitemOpen
  \bibfield  {author} {\bibinfo {author} {\bibfnamefont {M.~A.}\ \bibnamefont
  {Norcia}}, \bibinfo {author} {\bibfnamefont {R.~J.}\ \bibnamefont
  {Lewis-Swan}}, \bibinfo {author} {\bibfnamefont {J.~R.}\ \bibnamefont
  {Cline}}, \bibinfo {author} {\bibfnamefont {B.}~\bibnamefont {Zhu}}, \bibinfo
  {author} {\bibfnamefont {A.~M.}\ \bibnamefont {Rey}}, \ and\ \bibinfo
  {author} {\bibfnamefont {J.~K.}\ \bibnamefont {Thompson}},\ }\bibfield
  {title} {\emph {\bibinfo {title} {Cavity-mediated collective spin-exchange
  interactions in a strontium superradiant laser},\ }}\href {\doibase
  10.1126/science.aar3102} {\bibfield  {journal} {\bibinfo  {journal}
  {Science}\ }\textbf {\bibinfo {volume} {361}},\ \bibinfo {pages} {259}
  (\bibinfo {year} {2018})}\BibitemShut {NoStop}%
\bibitem [{\citenamefont {M{\o}lmer}\ and\ \citenamefont
  {S{\o}rensen}(1999)}]{molmer1999multiparticle}%
  \BibitemOpen
  \bibfield  {author} {\bibinfo {author} {\bibfnamefont {K.}~\bibnamefont
  {M{\o}lmer}}\ and\ \bibinfo {author} {\bibfnamefont {A.}~\bibnamefont
  {S{\o}rensen}},\ }\bibfield  {title} {\emph {\bibinfo {title} {Multiparticle
  entanglement of hot trapped ions},\ }}\href {\doibase
  10.1103/PhysRevLett.82.1835} {\bibfield  {journal} {\bibinfo  {journal}
  {Phys. Rev. Lett.}\ }\textbf {\bibinfo {volume} {82}},\ \bibinfo {pages}
  {1835} (\bibinfo {year} {1999})}\BibitemShut {NoStop}%
\bibitem [{\citenamefont {Britton}\ \emph {et~al.}(2012)\citenamefont
  {Britton}, \citenamefont {Sawyer}, \citenamefont {Keith}, \citenamefont
  {Wang}, \citenamefont {Freericks}, \citenamefont {Uys}, \citenamefont
  {Biercuk},\ and\ \citenamefont {Bollinger}}]{britton2012engineered}%
  \BibitemOpen
  \bibfield  {author} {\bibinfo {author} {\bibfnamefont {J.~W.}\ \bibnamefont
  {Britton}}, \bibinfo {author} {\bibfnamefont {B.~C.}\ \bibnamefont {Sawyer}},
  \bibinfo {author} {\bibfnamefont {A.~C.}\ \bibnamefont {Keith}}, \bibinfo
  {author} {\bibfnamefont {C.-C.~J.}\ \bibnamefont {Wang}}, \bibinfo {author}
  {\bibfnamefont {J.~K.}\ \bibnamefont {Freericks}}, \bibinfo {author}
  {\bibfnamefont {H.}~\bibnamefont {Uys}}, \bibinfo {author} {\bibfnamefont
  {M.~J.}\ \bibnamefont {Biercuk}}, \ and\ \bibinfo {author} {\bibfnamefont
  {J.~J.}\ \bibnamefont {Bollinger}},\ }\bibfield  {title} {\emph {\bibinfo
  {title} {Engineered two-dimensional ising interactions in a trapped-ion
  quantum simulator with hundreds of spins},\ }}\href {\doibase
  10.1038/nature10981} {\bibfield  {journal} {\bibinfo  {journal} {Nature}\
  }\textbf {\bibinfo {volume} {484}},\ \bibinfo {pages} {489} (\bibinfo {year}
  {2012})}\BibitemShut {NoStop}%
\bibitem [{\citenamefont {Bohnet}\ \emph {et~al.}(2016)\citenamefont {Bohnet},
  \citenamefont {Sawyer}, \citenamefont {Britton}, \citenamefont {Wall},
  \citenamefont {Rey}, \citenamefont {Foss-Feig},\ and\ \citenamefont
  {Bollinger}}]{bohnet2016quantum}%
  \BibitemOpen
  \bibfield  {author} {\bibinfo {author} {\bibfnamefont {J.~G.}\ \bibnamefont
  {Bohnet}}, \bibinfo {author} {\bibfnamefont {B.~C.}\ \bibnamefont {Sawyer}},
  \bibinfo {author} {\bibfnamefont {J.~W.}\ \bibnamefont {Britton}}, \bibinfo
  {author} {\bibfnamefont {M.~L.}\ \bibnamefont {Wall}}, \bibinfo {author}
  {\bibfnamefont {A.~M.}\ \bibnamefont {Rey}}, \bibinfo {author} {\bibfnamefont
  {M.}~\bibnamefont {Foss-Feig}}, \ and\ \bibinfo {author} {\bibfnamefont
  {J.~J.}\ \bibnamefont {Bollinger}},\ }\bibfield  {title} {\emph {\bibinfo
  {title} {Quantum spin dynamics and entanglement generation with hundreds of
  trapped ions},\ }}\href {\doibase https://doi.org/10.1126/science.aad9958}
  {\bibfield  {journal} {\bibinfo  {journal} {Science}\ }\textbf {\bibinfo
  {volume} {352}},\ \bibinfo {pages} {1297} (\bibinfo {year}
  {2016})}\BibitemShut {NoStop}%
\bibitem [{\citenamefont {S{\o}rensen}\ and\ \citenamefont
  {M{\o}lmer}(1999)}]{sorensen1999spin}%
  \BibitemOpen
  \bibfield  {author} {\bibinfo {author} {\bibfnamefont {A.}~\bibnamefont
  {S{\o}rensen}}\ and\ \bibinfo {author} {\bibfnamefont {K.}~\bibnamefont
  {M{\o}lmer}},\ }\bibfield  {title} {\emph {\bibinfo {title} {Spin-spin
  interaction and spin squeezing in an optical lattice},\ }}\href {\doibase
  10.1103/PhysRevLett.83.2274} {\bibfield  {journal} {\bibinfo  {journal}
  {Phys. Rev. Lett.}\ }\textbf {\bibinfo {volume} {83}},\ \bibinfo {pages}
  {2274} (\bibinfo {year} {1999})}\BibitemShut {NoStop}%
\bibitem [{\citenamefont {He}\ \emph {et~al.}(2019)\citenamefont {He},
  \citenamefont {Perlin}, \citenamefont {Muleady}, \citenamefont {Lewis-Swan},
  \citenamefont {Hutson}, \citenamefont {Ye},\ and\ \citenamefont
  {Rey}}]{he2019engineering}%
  \BibitemOpen
  \bibfield  {author} {\bibinfo {author} {\bibfnamefont {P.}~\bibnamefont
  {He}}, \bibinfo {author} {\bibfnamefont {M.~A.}\ \bibnamefont {Perlin}},
  \bibinfo {author} {\bibfnamefont {S.~R.}\ \bibnamefont {Muleady}}, \bibinfo
  {author} {\bibfnamefont {R.~J.}\ \bibnamefont {Lewis-Swan}}, \bibinfo
  {author} {\bibfnamefont {R.~B.}\ \bibnamefont {Hutson}}, \bibinfo {author}
  {\bibfnamefont {J.}~\bibnamefont {Ye}}, \ and\ \bibinfo {author}
  {\bibfnamefont {A.~M.}\ \bibnamefont {Rey}},\ }\bibfield  {title} {\emph
  {\bibinfo {title} {Engineering spin squeezing in a 3d optical lattice with
  interacting spin-orbit-coupled fermions},\ }}\href {\doibase
  10.1103/PhysRevResearch.1.033075} {\bibfield  {journal} {\bibinfo  {journal}
  {Phys. Rev. Res.}\ }\textbf {\bibinfo {volume} {1}},\ \bibinfo {pages}
  {033075} (\bibinfo {year} {2019})}\BibitemShut {NoStop}%
\bibitem [{\citenamefont {Hern{\'a}ndez~Yanes}\ \emph
  {et~al.}(2022)\citenamefont {Hern{\'a}ndez~Yanes}, \citenamefont
  {P{\l}odzie{\'n}}, \citenamefont {Mackoit~Sinkevi{\v{c}}ien{\.e}},
  \citenamefont {{\v{Z}}labys}, \citenamefont {Juzeli{\=u}nas},\ and\
  \citenamefont {Witkowska}}]{hernandez2022one}%
  \BibitemOpen
  \bibfield  {author} {\bibinfo {author} {\bibfnamefont {T.}~\bibnamefont
  {Hern{\'a}ndez~Yanes}}, \bibinfo {author} {\bibfnamefont {M.}~\bibnamefont
  {P{\l}odzie{\'n}}}, \bibinfo {author} {\bibfnamefont {M.}~\bibnamefont
  {Mackoit~Sinkevi{\v{c}}ien{\.e}}}, \bibinfo {author} {\bibfnamefont
  {G.}~\bibnamefont {{\v{Z}}labys}}, \bibinfo {author} {\bibfnamefont
  {G.}~\bibnamefont {Juzeli{\=u}nas}}, \ and\ \bibinfo {author} {\bibfnamefont
  {E.}~\bibnamefont {Witkowska}},\ }\bibfield  {title} {\emph {\bibinfo {title}
  {One-and two-axis squeezing via laser coupling in an atomic {Fermi-Hubbard}
  model},\ }}\href {\doibase 10.1103/PhysRevLett.129.090403} {\bibfield
  {journal} {\bibinfo  {journal} {Phys. Rev. Lett.}\ }\textbf {\bibinfo
  {volume} {129}},\ \bibinfo {pages} {090403} (\bibinfo {year}
  {2022})}\BibitemShut {NoStop}%
\bibitem [{\citenamefont {Boixo}\ \emph {et~al.}(2007)\citenamefont {Boixo},
  \citenamefont {Flammia}, \citenamefont {Caves},\ and\ \citenamefont
  {Geremia}}]{boixo2007generalized}%
  \BibitemOpen
  \bibfield  {author} {\bibinfo {author} {\bibfnamefont {S.}~\bibnamefont
  {Boixo}}, \bibinfo {author} {\bibfnamefont {S.~T.}\ \bibnamefont {Flammia}},
  \bibinfo {author} {\bibfnamefont {C.~M.}\ \bibnamefont {Caves}}, \ and\
  \bibinfo {author} {\bibfnamefont {M.}~\bibnamefont {Geremia}},\ }\bibfield
  {title} {\emph {\bibinfo {title} {Generalized limits for single-parameter
  quantum estimation},\ }}\href {\doibase 10.1103/PhysRevLett.98.090401}
  {\bibfield  {journal} {\bibinfo  {journal} {Phys. Rev. Lett.}\ }\textbf
  {\bibinfo {volume} {98}},\ \bibinfo {pages} {090401} (\bibinfo {year}
  {2007})}\BibitemShut {NoStop}%
\bibitem [{\citenamefont {Xia}\ \emph {et~al.}(2024)\citenamefont {Xia},
  \citenamefont {Huang}, \citenamefont {Li}, \citenamefont {Luo},\ and\
  \citenamefont {Zeng}}]{xia2023nanoradian}%
  \BibitemOpen
  \bibfield  {author} {\bibinfo {author} {\bibfnamefont {B.}~\bibnamefont
  {Xia}}, \bibinfo {author} {\bibfnamefont {J.}~\bibnamefont {Huang}}, \bibinfo
  {author} {\bibfnamefont {H.}~\bibnamefont {Li}}, \bibinfo {author}
  {\bibfnamefont {Z.}~\bibnamefont {Luo}}, \ and\ \bibinfo {author}
  {\bibfnamefont {G.}~\bibnamefont {Zeng}},\ }\bibfield  {title} {\emph
  {\bibinfo {title} {Nanoradian-scale precision in light rotation measurement
  via indefinite quantum dynamics},\ }}\href {\doibase 10.1126/sciadv.adm8524}
  {\bibfield  {journal} {\bibinfo  {journal} {Sci. Adv.}\ }\textbf {\bibinfo
  {volume} {10}},\ \bibinfo {pages} {eadm8524} (\bibinfo {year}
  {2024})}\BibitemShut {NoStop}%
\bibitem [{\citenamefont {Yang}\ \emph {et~al.}(2022)\citenamefont {Yang},
  \citenamefont {Pang}, \citenamefont {Chen}, \citenamefont {Jordan},\ and\
  \citenamefont {Del~Campo}}]{yang2022variational}%
  \BibitemOpen
  \bibfield  {author} {\bibinfo {author} {\bibfnamefont {J.}~\bibnamefont
  {Yang}}, \bibinfo {author} {\bibfnamefont {S.}~\bibnamefont {Pang}}, \bibinfo
  {author} {\bibfnamefont {Z.}~\bibnamefont {Chen}}, \bibinfo {author}
  {\bibfnamefont {A.~N.}\ \bibnamefont {Jordan}}, \ and\ \bibinfo {author}
  {\bibfnamefont {A.}~\bibnamefont {Del~Campo}},\ }\bibfield  {title} {\emph
  {\bibinfo {title} {Variational principle for optimal quantum controls in
  quantum metrology},\ }}\href {\doibase 10.1103/PhysRevLett.128.160505}
  {\bibfield  {journal} {\bibinfo  {journal} {Phys. Rev. Lett.}\ }\textbf
  {\bibinfo {volume} {128}},\ \bibinfo {pages} {160505} (\bibinfo {year}
  {2022})}\BibitemShut {NoStop}%
\bibitem [{\citenamefont {Luo}\ and\ \citenamefont {Yu}(2023)}]{luo2023time}%
  \BibitemOpen
  \bibfield  {author} {\bibinfo {author} {\bibfnamefont {D.-W.}\ \bibnamefont
  {Luo}}\ and\ \bibinfo {author} {\bibfnamefont {T.}~\bibnamefont {Yu}},\
  }\bibfield  {title} {\emph {\bibinfo {title} {Time-reversal assisted quantum
  metrology with an optimal control},\ }}\href
  {https://doi.org/10.48550/arXiv.2312.14443} {\bibfield  {journal} {\bibinfo
  {journal} {arXiv:2312.14443}\ } (\bibinfo {year} {2023})}\BibitemShut
  {NoStop}%
\bibitem [{\citenamefont {Fan}\ and\ \citenamefont
  {Pang}(2024)}]{fan2024achieving}%
  \BibitemOpen
  \bibfield  {author} {\bibinfo {author} {\bibfnamefont {J.}~\bibnamefont
  {Fan}}\ and\ \bibinfo {author} {\bibfnamefont {S.}~\bibnamefont {Pang}},\
  }\bibfield  {title} {\emph {\bibinfo {title} {Achieving {H}eisenberg scaling
  by probe-ancilla interaction in quantum metrology},\ }}\href {\doibase
  10.1103/PhysRevA.110.062406} {\bibfield  {journal} {\bibinfo  {journal}
  {Phys. Rev. A}\ }\textbf {\bibinfo {volume} {110}},\ \bibinfo {pages}
  {062406} (\bibinfo {year} {2024})}\BibitemShut {NoStop}%
\bibitem [{\citenamefont {Chen}\ and\ \citenamefont
  {Jing}(2024)}]{chen2024qubit}%
  \BibitemOpen
  \bibfield  {author} {\bibinfo {author} {\bibfnamefont {P.}~\bibnamefont
  {Chen}}\ and\ \bibinfo {author} {\bibfnamefont {J.}~\bibnamefont {Jing}},\
  }\bibfield  {title} {\emph {\bibinfo {title} {Qubit-assisted quantum
  metrology under a time-reversal strategy},\ }}\href {\doibase
  10.1103/PhysRevA.110.062425} {\bibfield  {journal} {\bibinfo  {journal}
  {Phys. Rev. A}\ }\textbf {\bibinfo {volume} {110}},\ \bibinfo {pages}
  {062425} (\bibinfo {year} {2024})}\BibitemShut {NoStop}%
\bibitem [{\citenamefont {Demkowicz-Dobrza\ifmmode~\acute{n}\else
  \'{n}\fi{}ski}\ and\ \citenamefont {Maccone}(2014)}]{demkowicz2014using}%
  \BibitemOpen
  \bibfield  {author} {\bibinfo {author} {\bibfnamefont {R.}~\bibnamefont
  {Demkowicz-Dobrza\ifmmode~\acute{n}\else \'{n}\fi{}ski}}\ and\ \bibinfo
  {author} {\bibfnamefont {L.}~\bibnamefont {Maccone}},\ }\bibfield  {title}
  {\emph {\bibinfo {title} {Using entanglement against noise in quantum
  metrology},\ }}\href {\doibase 10.1103/PhysRevLett.113.250801} {\bibfield
  {journal} {\bibinfo  {journal} {Phys. Rev. Lett.}\ }\textbf {\bibinfo
  {volume} {113}},\ \bibinfo {pages} {250801} (\bibinfo {year}
  {2014})}\BibitemShut {NoStop}%
\bibitem [{\citenamefont {He}\ \emph {et~al.}(2021)\citenamefont {He},
  \citenamefont {Guang}, \citenamefont {Li}, \citenamefont {Deng},
  \citenamefont {Zhang}, \citenamefont {Zhao}, \citenamefont {Deng},\ and\
  \citenamefont {Ai}}]{he2021quantum}%
  \BibitemOpen
  \bibfield  {author} {\bibinfo {author} {\bibfnamefont {W.-T.}\ \bibnamefont
  {He}}, \bibinfo {author} {\bibfnamefont {H.-Y.}\ \bibnamefont {Guang}},
  \bibinfo {author} {\bibfnamefont {Z.-Y.}\ \bibnamefont {Li}}, \bibinfo
  {author} {\bibfnamefont {R.-Q.}\ \bibnamefont {Deng}}, \bibinfo {author}
  {\bibfnamefont {N.-N.}\ \bibnamefont {Zhang}}, \bibinfo {author}
  {\bibfnamefont {J.-X.}\ \bibnamefont {Zhao}}, \bibinfo {author}
  {\bibfnamefont {F.-G.}\ \bibnamefont {Deng}}, \ and\ \bibinfo {author}
  {\bibfnamefont {Q.}~\bibnamefont {Ai}},\ }\bibfield  {title} {\emph {\bibinfo
  {title} {Quantum metrology with one auxiliary particle in a correlated bath
  and its quantum simulation},\ }}\href {\doibase 10.1103/PhysRevA.104.062429}
  {\bibfield  {journal} {\bibinfo  {journal} {Phys. Rev. A}\ }\textbf {\bibinfo
  {volume} {104}},\ \bibinfo {pages} {062429} (\bibinfo {year}
  {2021})}\BibitemShut {NoStop}%
\bibitem [{\citenamefont {Pezz\`e}\ \emph {et~al.}(2017)\citenamefont
  {Pezz\`e}, \citenamefont {Ciampini}, \citenamefont {Spagnolo}, \citenamefont
  {Humphreys}, \citenamefont {Datta}, \citenamefont {Walmsley}, \citenamefont
  {Barbieri}, \citenamefont {Sciarrino},\ and\ \citenamefont
  {Smerzi}}]{pezze2017optimal}%
  \BibitemOpen
  \bibfield  {author} {\bibinfo {author} {\bibfnamefont {L.}~\bibnamefont
  {Pezz\`e}}, \bibinfo {author} {\bibfnamefont {M.~A.}\ \bibnamefont
  {Ciampini}}, \bibinfo {author} {\bibfnamefont {N.}~\bibnamefont {Spagnolo}},
  \bibinfo {author} {\bibfnamefont {P.~C.}\ \bibnamefont {Humphreys}}, \bibinfo
  {author} {\bibfnamefont {A.}~\bibnamefont {Datta}}, \bibinfo {author}
  {\bibfnamefont {I.~A.}\ \bibnamefont {Walmsley}}, \bibinfo {author}
  {\bibfnamefont {M.}~\bibnamefont {Barbieri}}, \bibinfo {author}
  {\bibfnamefont {F.}~\bibnamefont {Sciarrino}}, \ and\ \bibinfo {author}
  {\bibfnamefont {A.}~\bibnamefont {Smerzi}},\ }\bibfield  {title} {\emph
  {\bibinfo {title} {Optimal measurements for simultaneous quantum estimation
  of multiple phases},\ }}\href {\doibase 10.1103/PhysRevLett.119.130504}
  {\bibfield  {journal} {\bibinfo  {journal} {Phys. Rev. Lett.}\ }\textbf
  {\bibinfo {volume} {119}},\ \bibinfo {pages} {130504} (\bibinfo {year}
  {2017})}\BibitemShut {NoStop}%
\bibitem [{\citenamefont {Yurke}\ \emph {et~al.}(1986)\citenamefont {Yurke},
  \citenamefont {McCall},\ and\ \citenamefont {Klauder}}]{yurke19862}%
  \BibitemOpen
  \bibfield  {author} {\bibinfo {author} {\bibfnamefont {B.}~\bibnamefont
  {Yurke}}, \bibinfo {author} {\bibfnamefont {S.~L.}\ \bibnamefont {McCall}}, \
  and\ \bibinfo {author} {\bibfnamefont {J.~R.}\ \bibnamefont {Klauder}},\
  }\bibfield  {title} {\emph {\bibinfo {title} {{SU}(2) and {SU}(1,1)
  interferometers},\ }}\href {\doibase 10.1103/PhysRevA.33.4033} {\bibfield
  {journal} {\bibinfo  {journal} {Phys. Rev. A}\ }\textbf {\bibinfo {volume}
  {33}},\ \bibinfo {pages} {4033} (\bibinfo {year} {1986})}\BibitemShut
  {NoStop}%
\bibitem [{\citenamefont {Chen}\ and\ \citenamefont
  {Jing}(2025)}]{ChenJing2025}%
  \BibitemOpen
  \bibfield  {author} {\bibinfo {author} {\bibfnamefont {P.}~\bibnamefont
  {Chen}}\ and\ \bibinfo {author} {\bibfnamefont {J.}~\bibnamefont {Jing}},\
  }\bibfield  {title} {\emph {\bibinfo {title} {Achieving the {H}eisenberg
  limit of metrology via measurement on an ancillary qubit},\ }}\href {\doibase
  10.1103/f79z-vjsb} {\bibfield  {journal} {\bibinfo  {journal} {Phys. Rev. A}\
  }\textbf {\bibinfo {volume} {112}},\ \bibinfo {pages} {032416} (\bibinfo
  {year} {2025})}\BibitemShut {NoStop}%
\bibitem [{\citenamefont {Barry}\ \emph {et~al.}(2020)\citenamefont {Barry},
  \citenamefont {Schloss}, \citenamefont {Bauch}, \citenamefont {Turner},
  \citenamefont {Hart}, \citenamefont {Pham},\ and\ \citenamefont
  {Walsworth}}]{barry2020sensitivity}%
  \BibitemOpen
  \bibfield  {author} {\bibinfo {author} {\bibfnamefont {J.~F.}\ \bibnamefont
  {Barry}}, \bibinfo {author} {\bibfnamefont {J.~M.}\ \bibnamefont {Schloss}},
  \bibinfo {author} {\bibfnamefont {E.}~\bibnamefont {Bauch}}, \bibinfo
  {author} {\bibfnamefont {M.~J.}\ \bibnamefont {Turner}}, \bibinfo {author}
  {\bibfnamefont {C.~A.}\ \bibnamefont {Hart}}, \bibinfo {author}
  {\bibfnamefont {L.~M.}\ \bibnamefont {Pham}}, \ and\ \bibinfo {author}
  {\bibfnamefont {R.~L.}\ \bibnamefont {Walsworth}},\ }\bibfield  {title}
  {\emph {\bibinfo {title} {Sensitivity optimization for {NV}-diamond
  magnetometry},\ }}\href {\doibase 10.1103/RevModPhys.92.015004} {\bibfield
  {journal} {\bibinfo  {journal} {Rev. Mod. Phys.}\ }\textbf {\bibinfo {volume}
  {92}},\ \bibinfo {pages} {015004} (\bibinfo {year} {2020})}\BibitemShut
  {NoStop}%
\bibitem [{\citenamefont {He}\ \emph {et~al.}(1993)\citenamefont {He},
  \citenamefont {Manson},\ and\ \citenamefont {Fisk}}]{he1993paramagnetic}%
  \BibitemOpen
  \bibfield  {author} {\bibinfo {author} {\bibfnamefont {X.-F.}\ \bibnamefont
  {He}}, \bibinfo {author} {\bibfnamefont {N.~B.}\ \bibnamefont {Manson}}, \
  and\ \bibinfo {author} {\bibfnamefont {P.~T.}\ \bibnamefont {Fisk}},\
  }\bibfield  {title} {\emph {\bibinfo {title} {Paramagnetic resonance of
  photoexcited {N-V} defects in diamond. {II}. hyperfine interaction with the
  {$^{14}N$} nucleus},\ }}\href {\doibase 10.1103/PhysRevB.47.8816} {\bibfield
  {journal} {\bibinfo  {journal} {Physical Review B}\ }\textbf {\bibinfo
  {volume} {47}},\ \bibinfo {pages} {8816} (\bibinfo {year}
  {1993})}\BibitemShut {NoStop}%
\bibitem [{\citenamefont {Yao}\ \emph {et~al.}(2006)\citenamefont {Yao},
  \citenamefont {Liu},\ and\ \citenamefont {Sham}}]{yao2006theory}%
  \BibitemOpen
  \bibfield  {author} {\bibinfo {author} {\bibfnamefont {W.}~\bibnamefont
  {Yao}}, \bibinfo {author} {\bibfnamefont {R.-B.}\ \bibnamefont {Liu}}, \ and\
  \bibinfo {author} {\bibfnamefont {L.~J.}\ \bibnamefont {Sham}},\ }\bibfield
  {title} {\emph {\bibinfo {title} {Theory of electron spin decoherence by
  interacting nuclear spins in a quantum dot},\ }}\href {\doibase
  10.1103/PhysRevB.74.195301} {\bibfield  {journal} {\bibinfo  {journal} {Phys.
  Rev. B}\ }\textbf {\bibinfo {volume} {74}},\ \bibinfo {pages} {195301}
  (\bibinfo {year} {2006})}\BibitemShut {NoStop}%
\bibitem [{\citenamefont {Liu}\ \emph {et~al.}(2007)\citenamefont {Liu},
  \citenamefont {Yao},\ and\ \citenamefont {Sham}}]{liu2007control}%
  \BibitemOpen
  \bibfield  {author} {\bibinfo {author} {\bibfnamefont {R.-B.}\ \bibnamefont
  {Liu}}, \bibinfo {author} {\bibfnamefont {W.}~\bibnamefont {Yao}}, \ and\
  \bibinfo {author} {\bibfnamefont {L.}~\bibnamefont {Sham}},\ }\bibfield
  {title} {\emph {\bibinfo {title} {Control of electron spin decoherence caused
  by electron--nuclear spin dynamics in a quantum dot},\ }}\href {\doibase
  10.1088/1367-2630/9/7/226} {\bibfield  {journal} {\bibinfo  {journal} {New J.
  Phys.}\ }\textbf {\bibinfo {volume} {9}},\ \bibinfo {pages} {226} (\bibinfo
  {year} {2007})}\BibitemShut {NoStop}%
\bibitem [{\citenamefont {Meyer}\ \emph {et~al.}(2001)\citenamefont {Meyer},
  \citenamefont {Rowe}, \citenamefont {Kielpinski}, \citenamefont {Sackett},
  \citenamefont {Itano}, \citenamefont {Monroe},\ and\ \citenamefont
  {Wineland}}]{meyer2001experimental}%
  \BibitemOpen
  \bibfield  {author} {\bibinfo {author} {\bibfnamefont {V.}~\bibnamefont
  {Meyer}}, \bibinfo {author} {\bibfnamefont {M.~A.}\ \bibnamefont {Rowe}},
  \bibinfo {author} {\bibfnamefont {D.}~\bibnamefont {Kielpinski}}, \bibinfo
  {author} {\bibfnamefont {C.~A.}\ \bibnamefont {Sackett}}, \bibinfo {author}
  {\bibfnamefont {W.~M.}\ \bibnamefont {Itano}}, \bibinfo {author}
  {\bibfnamefont {C.}~\bibnamefont {Monroe}}, \ and\ \bibinfo {author}
  {\bibfnamefont {D.~J.}\ \bibnamefont {Wineland}},\ }\bibfield  {title} {\emph
  {\bibinfo {title} {Experimental demonstration of entanglement-enhanced
  rotation angle estimation using trapped ions},\ }}\href {\doibase
  10.1103/PhysRevLett.86.5870} {\bibfield  {journal} {\bibinfo  {journal}
  {Phys. Rev. Lett.}\ }\textbf {\bibinfo {volume} {86}},\ \bibinfo {pages}
  {5870} (\bibinfo {year} {2001})}\BibitemShut {NoStop}%
\bibitem [{\citenamefont {Ockeloen}\ \emph {et~al.}(2013)\citenamefont
  {Ockeloen}, \citenamefont {Schmied}, \citenamefont {Riedel},\ and\
  \citenamefont {Treutlein}}]{ockeloen2013quantum}%
  \BibitemOpen
  \bibfield  {author} {\bibinfo {author} {\bibfnamefont {C.~F.}\ \bibnamefont
  {Ockeloen}}, \bibinfo {author} {\bibfnamefont {R.}~\bibnamefont {Schmied}},
  \bibinfo {author} {\bibfnamefont {M.~F.}\ \bibnamefont {Riedel}}, \ and\
  \bibinfo {author} {\bibfnamefont {P.}~\bibnamefont {Treutlein}},\ }\bibfield
  {title} {\emph {\bibinfo {title} {Quantum metrology with a scanning probe
  atom interferometer},\ }}\href {\doibase 10.1103/PhysRevLett.111.143001}
  {\bibfield  {journal} {\bibinfo  {journal} {Phys. Rev. Lett.}\ }\textbf
  {\bibinfo {volume} {111}},\ \bibinfo {pages} {143001} (\bibinfo {year}
  {2013})}\BibitemShut {NoStop}%
\bibitem [{\citenamefont {Xie}\ \emph {et~al.}(2021)\citenamefont {Xie},
  \citenamefont {Zhao}, \citenamefont {Kong}, \citenamefont {Ma}, \citenamefont
  {Wang}, \citenamefont {Ye}, \citenamefont {Yu}, \citenamefont {Yang},
  \citenamefont {Xu}, \citenamefont {Wang} \emph {et~al.}}]{xie2021beating}%
  \BibitemOpen
  \bibfield  {author} {\bibinfo {author} {\bibfnamefont {T.}~\bibnamefont
  {Xie}}, \bibinfo {author} {\bibfnamefont {Z.}~\bibnamefont {Zhao}}, \bibinfo
  {author} {\bibfnamefont {X.}~\bibnamefont {Kong}}, \bibinfo {author}
  {\bibfnamefont {W.}~\bibnamefont {Ma}}, \bibinfo {author} {\bibfnamefont
  {M.}~\bibnamefont {Wang}}, \bibinfo {author} {\bibfnamefont {X.}~\bibnamefont
  {Ye}}, \bibinfo {author} {\bibfnamefont {P.}~\bibnamefont {Yu}}, \bibinfo
  {author} {\bibfnamefont {Z.}~\bibnamefont {Yang}}, \bibinfo {author}
  {\bibfnamefont {S.}~\bibnamefont {Xu}}, \bibinfo {author} {\bibfnamefont
  {P.}~\bibnamefont {Wang}},  \emph {et~al.},\ }\bibfield  {title} {\emph
  {\bibinfo {title} {Beating the standard quantum limit under ambient
  conditions with solid-state spins},\ }}\href {\doibase
  10.1126/sciadv.abg9204} {\bibfield  {journal} {\bibinfo  {journal} {Sci.
  Adv.}\ }\textbf {\bibinfo {volume} {7}},\ \bibinfo {pages} {eabg9204}
  (\bibinfo {year} {2021})}\BibitemShut {NoStop}%
\bibitem [{\citenamefont {Pang}\ and\ \citenamefont
  {Brun}(2014)}]{pang2014quantum}%
  \BibitemOpen
  \bibfield  {author} {\bibinfo {author} {\bibfnamefont {S.}~\bibnamefont
  {Pang}}\ and\ \bibinfo {author} {\bibfnamefont {T.~A.}\ \bibnamefont
  {Brun}},\ }\bibfield  {title} {\emph {\bibinfo {title} {Quantum metrology for
  a general {Hamiltonian} parameter},\ }}\href {\doibase
  10.1103/PhysRevA.90.022117} {\bibfield  {journal} {\bibinfo  {journal} {Phys.
  Rev. A}\ }\textbf {\bibinfo {volume} {90}},\ \bibinfo {pages} {022117}
  (\bibinfo {year} {2014})}\BibitemShut {NoStop}%
\bibitem [{\citenamefont {Zhao}\ \emph {et~al.}(2020)\citenamefont {Zhao},
  \citenamefont {Yang},\ and\ \citenamefont {Chiribella}}]{zhao2020quantum}%
  \BibitemOpen
  \bibfield  {author} {\bibinfo {author} {\bibfnamefont {X.}~\bibnamefont
  {Zhao}}, \bibinfo {author} {\bibfnamefont {Y.}~\bibnamefont {Yang}}, \ and\
  \bibinfo {author} {\bibfnamefont {G.}~\bibnamefont {Chiribella}},\ }\bibfield
   {title} {\emph {\bibinfo {title} {Quantum metrology with indefinite causal
  order},\ }}\href {\doibase 10.1103/PhysRevLett.124.190503} {\bibfield
  {journal} {\bibinfo  {journal} {Phys. Rev. Lett.}\ }\textbf {\bibinfo
  {volume} {124}},\ \bibinfo {pages} {190503} (\bibinfo {year}
  {2020})}\BibitemShut {NoStop}%
\bibitem [{\citenamefont {Yin}\ \emph {et~al.}(2023)\citenamefont {Yin},
  \citenamefont {Zhao}, \citenamefont {Yang}, \citenamefont {Guo},
  \citenamefont {Zhang}, \citenamefont {Li}, \citenamefont {Han}, \citenamefont
  {Liu}, \citenamefont {Xu}, \citenamefont {Chiribella} \emph
  {et~al.}}]{yin2023experimental}%
  \BibitemOpen
  \bibfield  {author} {\bibinfo {author} {\bibfnamefont {P.}~\bibnamefont
  {Yin}}, \bibinfo {author} {\bibfnamefont {X.}~\bibnamefont {Zhao}}, \bibinfo
  {author} {\bibfnamefont {Y.}~\bibnamefont {Yang}}, \bibinfo {author}
  {\bibfnamefont {Y.}~\bibnamefont {Guo}}, \bibinfo {author} {\bibfnamefont
  {W.-H.}\ \bibnamefont {Zhang}}, \bibinfo {author} {\bibfnamefont {G.-C.}\
  \bibnamefont {Li}}, \bibinfo {author} {\bibfnamefont {Y.-J.}\ \bibnamefont
  {Han}}, \bibinfo {author} {\bibfnamefont {B.-H.}\ \bibnamefont {Liu}},
  \bibinfo {author} {\bibfnamefont {J.-S.}\ \bibnamefont {Xu}}, \bibinfo
  {author} {\bibfnamefont {G.}~\bibnamefont {Chiribella}},  \emph {et~al.},\
  }\bibfield  {title} {\emph {\bibinfo {title} {Experimental super-{H}eisenberg
  quantum metrology with indefinite gate order},\ }}\href {\doibase
  https://doi.org/10.1038/s41567-023-02046-y} {\bibfield  {journal} {\bibinfo
  {journal} {Nat. Phys.}\ }\textbf {\bibinfo {volume} {19}},\ \bibinfo {pages}
  {1122} (\bibinfo {year} {2023})}\BibitemShut {NoStop}%
\bibitem [{\citenamefont {Braunstein}\ and\ \citenamefont
  {Caves}(1994)}]{braunstein1994statistical}%
  \BibitemOpen
  \bibfield  {author} {\bibinfo {author} {\bibfnamefont {S.~L.}\ \bibnamefont
  {Braunstein}}\ and\ \bibinfo {author} {\bibfnamefont {C.~M.}\ \bibnamefont
  {Caves}},\ }\bibfield  {title} {\emph {\bibinfo {title} {Statistical distance
  and the geometry of quantum states},\ }}\href {\doibase
  10.1103/PhysRevLett.72.3439} {\bibfield  {journal} {\bibinfo  {journal}
  {Phys. Rev. Lett.}\ }\textbf {\bibinfo {volume} {72}},\ \bibinfo {pages}
  {3439} (\bibinfo {year} {1994})}\BibitemShut {NoStop}%
\bibitem [{\citenamefont {Demkowicz-Dobrza{\'n}ski}\ \emph
  {et~al.}(2015)\citenamefont {Demkowicz-Dobrza{\'n}ski}, \citenamefont
  {Jarzyna},\ and\ \citenamefont {Ko{\l}ody{\'n}ski}}]{demkowicz2015quantum}%
  \BibitemOpen
  \bibfield  {author} {\bibinfo {author} {\bibfnamefont {R.}~\bibnamefont
  {Demkowicz-Dobrza{\'n}ski}}, \bibinfo {author} {\bibfnamefont
  {M.}~\bibnamefont {Jarzyna}}, \ and\ \bibinfo {author} {\bibfnamefont
  {J.}~\bibnamefont {Ko{\l}ody{\'n}ski}},\ }\bibfield  {title} {\emph {\bibinfo
  {title} {Quantum limits in optical interferometry},\ }}\href {\doibase
  https://doi.org/10.1016/bs.po.2015.02.003} {\bibfield  {journal} {\bibinfo
  {journal} {Prog. Opt.}\ }\textbf {\bibinfo {volume} {60}},\ \bibinfo {pages}
  {345} (\bibinfo {year} {2015})}\BibitemShut {NoStop}%
\bibitem [{\citenamefont {Bradshaw}\ \emph {et~al.}(2018)\citenamefont
  {Bradshaw}, \citenamefont {Lam},\ and\ \citenamefont
  {Assad}}]{bradshaw2018ultimate}%
  \BibitemOpen
  \bibfield  {author} {\bibinfo {author} {\bibfnamefont {M.}~\bibnamefont
  {Bradshaw}}, \bibinfo {author} {\bibfnamefont {P.~K.}\ \bibnamefont {Lam}}, \
  and\ \bibinfo {author} {\bibfnamefont {S.~M.}\ \bibnamefont {Assad}},\
  }\bibfield  {title} {\emph {\bibinfo {title} {Ultimate precision of joint
  quadrature parameter estimation with a gaussian probe},\ }}\href {\doibase
  10.1103/PhysRevA.97.012106} {\bibfield  {journal} {\bibinfo  {journal} {Phys.
  Rev. A}\ }\textbf {\bibinfo {volume} {97}},\ \bibinfo {pages} {012106}
  (\bibinfo {year} {2018})}\BibitemShut {NoStop}%
\bibitem [{\citenamefont {Helstrom}(1969)}]{helstrom1969quantum}%
  \BibitemOpen
  \bibfield  {author} {\bibinfo {author} {\bibfnamefont {C.~W.}\ \bibnamefont
  {Helstrom}},\ }\bibfield  {title} {\emph {\bibinfo {title} {Quantum detection
  and estimation theory},\ }}\href {\doibase
  https://doi.org/10.1007/BF01007479} {\bibfield  {journal} {\bibinfo
  {journal} {J. Stat. Phys.}\ }\textbf {\bibinfo {volume} {1}},\ \bibinfo
  {pages} {231} (\bibinfo {year} {1969})}\BibitemShut {NoStop}%
\bibitem [{\citenamefont {Yuen}\ and\ \citenamefont
  {Chan}(1983)}]{yuen1983noise}%
  \BibitemOpen
  \bibfield  {author} {\bibinfo {author} {\bibfnamefont {H.~P.}\ \bibnamefont
  {Yuen}}\ and\ \bibinfo {author} {\bibfnamefont {V.~W.}\ \bibnamefont
  {Chan}},\ }\bibfield  {title} {\emph {\bibinfo {title} {Noise in homodyne and
  heterodyne detection},\ }}\href {\doibase 10.1364/OL.8.000177} {\bibfield
  {journal} {\bibinfo  {journal} {Opt. lett.}\ }\textbf {\bibinfo {volume}
  {8}},\ \bibinfo {pages} {177} (\bibinfo {year} {1983})}\BibitemShut {NoStop}%
\bibitem [{\citenamefont {Manceau}\ \emph {et~al.}(2017)\citenamefont
  {Manceau}, \citenamefont {Khalili},\ and\ \citenamefont
  {Chekhova}}]{manceau2017improving}%
  \BibitemOpen
  \bibfield  {author} {\bibinfo {author} {\bibfnamefont {M.}~\bibnamefont
  {Manceau}}, \bibinfo {author} {\bibfnamefont {F.}~\bibnamefont {Khalili}}, \
  and\ \bibinfo {author} {\bibfnamefont {M.}~\bibnamefont {Chekhova}},\
  }\bibfield  {title} {\emph {\bibinfo {title} {Improving the phase
  super-sensitivity of squeezing-assisted interferometers by squeeze factor
  unbalancing},\ }}\href {\doibase 10.1088/1367-2630/aa53d1} {\bibfield
  {journal} {\bibinfo  {journal} {New J. Phys.}\ }\textbf {\bibinfo {volume}
  {19}},\ \bibinfo {pages} {013014} (\bibinfo {year} {2017})}\BibitemShut
  {NoStop}%
\bibitem [{\citenamefont {Szigeti}\ \emph {et~al.}(2017)\citenamefont
  {Szigeti}, \citenamefont {Lewis-Swan},\ and\ \citenamefont
  {Haine}}]{szigeti2017pumped}%
  \BibitemOpen
  \bibfield  {author} {\bibinfo {author} {\bibfnamefont {S.~S.}\ \bibnamefont
  {Szigeti}}, \bibinfo {author} {\bibfnamefont {R.~J.}\ \bibnamefont
  {Lewis-Swan}}, \ and\ \bibinfo {author} {\bibfnamefont {S.~A.}\ \bibnamefont
  {Haine}},\ }\bibfield  {title} {\emph {\bibinfo {title} {Pumped-up {SU} (1,
  1) interferometry},\ }}\href {\doibase 10.1103/PhysRevLett.118.150401}
  {\bibfield  {journal} {\bibinfo  {journal} {Phys. Rev. Lett.}\ }\textbf
  {\bibinfo {volume} {118}},\ \bibinfo {pages} {150401} (\bibinfo {year}
  {2017})}\BibitemShut {NoStop}%
\bibitem [{\citenamefont {Ataman}\ \emph {et~al.}(2018)\citenamefont {Ataman},
  \citenamefont {Preda},\ and\ \citenamefont {Ionicioiu}}]{ataman2018phase}%
  \BibitemOpen
  \bibfield  {author} {\bibinfo {author} {\bibfnamefont {S.}~\bibnamefont
  {Ataman}}, \bibinfo {author} {\bibfnamefont {A.}~\bibnamefont {Preda}}, \
  and\ \bibinfo {author} {\bibfnamefont {R.}~\bibnamefont {Ionicioiu}},\
  }\bibfield  {title} {\emph {\bibinfo {title} {Phase sensitivity of a
  mach-zehnder interferometer with single-intensity and difference-intensity
  detection},\ }}\href {\doibase 10.1103/PhysRevA.98.043856} {\bibfield
  {journal} {\bibinfo  {journal} {Phys. Rev. A}\ }\textbf {\bibinfo {volume}
  {98}},\ \bibinfo {pages} {043856} (\bibinfo {year} {2018})}\BibitemShut
  {NoStop}%
\bibitem [{\citenamefont {Bollinger}\ \emph {et~al.}(1996)\citenamefont
  {Bollinger}, \citenamefont {Itano}, \citenamefont {Wineland},\ and\
  \citenamefont {Heinzen}}]{bollinger1996optimal}%
  \BibitemOpen
  \bibfield  {author} {\bibinfo {author} {\bibfnamefont {J.~J.}\ \bibnamefont
  {Bollinger}}, \bibinfo {author} {\bibfnamefont {W.~M.}\ \bibnamefont
  {Itano}}, \bibinfo {author} {\bibfnamefont {D.~J.}\ \bibnamefont {Wineland}},
  \ and\ \bibinfo {author} {\bibfnamefont {D.~J.}\ \bibnamefont {Heinzen}},\
  }\bibfield  {title} {\emph {\bibinfo {title} {Optimal frequency measurements
  with maximally correlated states},\ }}\href {\doibase
  https://doi.org/10.1103/PhysRevA.54.R4649} {\bibfield  {journal} {\bibinfo
  {journal} {Phys. Rev. A}\ }\textbf {\bibinfo {volume} {54}},\ \bibinfo
  {pages} {R4649} (\bibinfo {year} {1996})}\BibitemShut {NoStop}%
\bibitem [{\citenamefont {Anisimov}\ \emph {et~al.}(2010)\citenamefont
  {Anisimov}, \citenamefont {Raterman}, \citenamefont {Chiruvelli},
  \citenamefont {Plick}, \citenamefont {Huver}, \citenamefont {Lee},\ and\
  \citenamefont {Dowling}}]{anisimov2010quantum}%
  \BibitemOpen
  \bibfield  {author} {\bibinfo {author} {\bibfnamefont {P.~M.}\ \bibnamefont
  {Anisimov}}, \bibinfo {author} {\bibfnamefont {G.~M.}\ \bibnamefont
  {Raterman}}, \bibinfo {author} {\bibfnamefont {A.}~\bibnamefont
  {Chiruvelli}}, \bibinfo {author} {\bibfnamefont {W.~N.}\ \bibnamefont
  {Plick}}, \bibinfo {author} {\bibfnamefont {S.~D.}\ \bibnamefont {Huver}},
  \bibinfo {author} {\bibfnamefont {H.}~\bibnamefont {Lee}}, \ and\ \bibinfo
  {author} {\bibfnamefont {J.~P.}\ \bibnamefont {Dowling}},\ }\bibfield
  {title} {\emph {\bibinfo {title} {Quantum metrology with two-mode squeezed
  vacuum: Parity detection beats the {H}eisenberg limit},\ }}\href {\doibase
  10.1103/PhysRevLett.104.103602} {\bibfield  {journal} {\bibinfo  {journal}
  {Phys. Rev. Lett.}\ }\textbf {\bibinfo {volume} {104}},\ \bibinfo {pages}
  {103602} (\bibinfo {year} {2010})}\BibitemShut {NoStop}%
\bibitem [{\citenamefont {Kielinski}\ \emph {et~al.}(2024)\citenamefont
  {Kielinski}, \citenamefont {Schmidt},\ and\ \citenamefont
  {Hammerer}}]{kielinski2024ghz}%
  \BibitemOpen
  \bibfield  {author} {\bibinfo {author} {\bibfnamefont {T.}~\bibnamefont
  {Kielinski}}, \bibinfo {author} {\bibfnamefont {P.~O.}\ \bibnamefont
  {Schmidt}}, \ and\ \bibinfo {author} {\bibfnamefont {K.}~\bibnamefont
  {Hammerer}},\ }\bibfield  {title} {\emph {\bibinfo {title} {{GHZ} protocols
  enhance frequency metrology despite spontaneous decay},\ }}\href {\doibase
  10.1126/sciadv.adr1439} {\bibfield  {journal} {\bibinfo  {journal} {Sci.
  Adv.}\ }\textbf {\bibinfo {volume} {10}},\ \bibinfo {pages} {eadr1439}
  (\bibinfo {year} {2024})}\BibitemShut {NoStop}%
\bibitem [{\citenamefont {Deng}\ \emph {et~al.}(2024)\citenamefont {Deng},
  \citenamefont {Li}, \citenamefont {Chen}, \citenamefont {Ni}, \citenamefont
  {Cai}, \citenamefont {Mai}, \citenamefont {Zhang}, \citenamefont {Zheng},
  \citenamefont {Yu}, \citenamefont {Zou} \emph {et~al.}}]{deng2024quantum}%
  \BibitemOpen
  \bibfield  {author} {\bibinfo {author} {\bibfnamefont {X.}~\bibnamefont
  {Deng}}, \bibinfo {author} {\bibfnamefont {S.}~\bibnamefont {Li}}, \bibinfo
  {author} {\bibfnamefont {Z.-J.}\ \bibnamefont {Chen}}, \bibinfo {author}
  {\bibfnamefont {Z.}~\bibnamefont {Ni}}, \bibinfo {author} {\bibfnamefont
  {Y.}~\bibnamefont {Cai}}, \bibinfo {author} {\bibfnamefont {J.}~\bibnamefont
  {Mai}}, \bibinfo {author} {\bibfnamefont {L.}~\bibnamefont {Zhang}}, \bibinfo
  {author} {\bibfnamefont {P.}~\bibnamefont {Zheng}}, \bibinfo {author}
  {\bibfnamefont {H.}~\bibnamefont {Yu}}, \bibinfo {author} {\bibfnamefont
  {C.-L.}\ \bibnamefont {Zou}},  \emph {et~al.},\ }\bibfield  {title} {\emph
  {\bibinfo {title} {Quantum-enhanced metrology with large {Fock} states},\
  }}\href {\doibase 10.1038/s41567-024-02619-5} {\bibfield  {journal} {\bibinfo
   {journal} {Nat. Phys.}\ }\textbf {\bibinfo {volume} {20}},\ \bibinfo {pages}
  {1874} (\bibinfo {year} {2024})}\BibitemShut {NoStop}%
\end{thebibliography}%

\end{document}